\documentclass[lettersize,journal]{IEEEtran}
\usepackage{amsmath,amsfonts}
\usepackage{algorithmic}
\usepackage{algorithm}
\usepackage{array}
\usepackage[caption=false,font=footnotesize,labelfont=rm,textfont=rm]{subfig}
\usepackage{textcomp}
\usepackage{stfloats}
\usepackage{url}
\usepackage{verbatim}
\usepackage{graphicx}
\usepackage{cite}
\usepackage{cite}
\usepackage{graphicx}
\usepackage{subfig}
\usepackage{subcaption}
\usepackage{amsmath}
\usepackage{amssymb}
\usepackage{balance}
\usepackage{booktabs,multirow}
\usepackage{graphicx}
\usepackage{graphicx}
\usepackage[most]{tcolorbox}
\usepackage[table]{xcolor}
\usepackage{makecell}
\usepackage{array}
\usepackage{makecell} 
\usepackage{colortbl} 
\usepackage{booktabs}

\newcolumntype{C}{>{\centering\arraybackslash}p}

\tcbset{
    answerbox/.style={
    enhanced,
    colback=gray!10,           
    colframe=white,            
    boxrule=0pt,               
    left=15pt,                 
    right=15pt,                
    top=6pt,
    bottom=6pt,
    before skip=8pt,           
    after skip=8pt,            
    overlay unbroken and first={%
      \draw[line width=2pt, black] (frame.north west) -- (frame.south west); 
      \draw[line width=2pt, black] (frame.north east) -- (frame.south east); 
        },
    }
}

\begin{document}
\title{From LLMs to Agents: A Comparative Evaluation of LLMs and LLM-based Agents in Security Patch Detection}
\author{Junxiao Han, Zheng Yu, Lingfeng Bao, Jiakun Liu, Yao Wan, Jianwei Yin, Shuiguang Deng, and Song Han 
\IEEEcompsocitemizethanks{\IEEEcompsocthanksitem  Junxiao Han, Zheng Yu, and Song Han are with the School of Computer and Computing Science, Hangzhou City University, Hangzhou 310015, China. E-mail: hanjx@hzcu.edu.cn, yuzheng.hzcu@gmail.com, and hans@hzcu.edu.cn
\IEEEcompsocthanksitem Lingfeng Bao is with the State Key Laboratory of Blockchain and Data Security, Zhejiang University, Hangzhou 310027, China. E-mail: lingfengbao@zju.edu.cn 
\IEEEcompsocthanksitem Jiakun Liu is with the Faculty of Computing, Harbin Institute of Technology, Harbin 150001, China. E-mail: jiakunliu@hit.edu.cn
\IEEEcompsocthanksitem Yao Wan is with the College of Computer Science and Technology, Huazhong University of Science and Technology, Wuhan, China. E-mail: wanyao@hust.edu.cn
\IEEEcompsocthanksitem Jianwei Yin is with the College of Software Technology, Zhejiang University, Ningbo 315100, China. E-mail: zjuyjw@cs.zju.edu.cn
\IEEEcompsocthanksitem Shuiguang Deng is with the College of Computer Science and Technology, Zhejiang University, Hangzhou 310027, China. E-mail: dengsg@zju.edu.cn
\IEEEcompsocthanksitem Song Han is the corresponding author.
}}


\maketitle

\begin{abstract}
The widespread adoption of open-source software (OSS) has accelerated software innovation but also increased security risks due to the rapid propagation of vulnerabilities and silent patch releases. In recent years, large language models (LLMs) and LLM-based agents have demonstrated remarkable capabilities in various software engineering (SE) tasks, enabling them to effectively address software security challenges such as vulnerability detection. However, systematic evaluation of the capabilities of LLMs and LLM-based agents in security patch detection remains limited. To bridge this gap, we conduct a comprehensive evaluation of the performance of LLMs and LLM-based agents for security patch detection. Specifically, we investigate three methods: Plain LLM (a single LLM with a system prompt), Data-Aug LLM (data augmentation based on the Plain LLM), and the ReAct Agent (leveraging the thought-action-observation mechanism). We also evaluate the performance of both commercial and open-source LLMs under these methods and compare these results with those of existing baselines. Furthermore, we analyze the detection performance of these methods across various vulnerability types, and examine the impact of different prompting strategies and context window sizes on the results. Our findings reveal that the Data-Aug LLM achieves the best overall performance, whereas the ReAct Agent demonstrates the lowest false positive rate (FPR). Although baseline methods exhibit strong accuracy, their false positive rates are significantly higher. In contrast, our evaluated methods achieve comparable accuracy while substantially reducing the FPR. These findings provide valuable insights into the practical applications of LLMs and LLM-based agents in security patch detection, highlighting their advantage in maintaining robust performance while minimizing false positive rates.

\end{abstract}

\section{Introduction} \label{intro}.
\IEEEPARstart {A}{lthough} the open-source software (OSS) movement has greatly advanced software development, the rapid diffusion of vulnerabilities in OSS has substantially increased security risks. According to the Black Duck 2025 report \cite{blackduck}, 97\% of codebases depend on OSS, with 86\% containing at least one known open-source vulnerability, and 81\% involving issues of high or critical severity. Moreover, 91\% of these codebases include components that are outdated by ten or more versions \cite{Synopsys2024OSSRA}. This highlights the urgent need for timely detection of software security patches to mitigate potential attacks \cite{wang2019detecting}.

However, the management of security patches is often subjective \cite{wang2019detecting,tamjidyamcholo2022subjectivity}, as software vendors may release security updates silently, without sufficient public disclosure \cite{li2017large}. Given that commits frequently include a wide range of changes, such as introducing new features, optimizing performance, updating versions, or releasing security patches, this practice of silent patch releases often complicates the processes of vulnerability detection and remediation. Security patches are frequently overwhelmed by the increasing volume of commits and patches \cite{mirhosseini2017can}, leading to delays in software updates and vulnerability reporting. Existing research indicates that over 82\% of user-submitted software vulnerability reports are filed more than 30 days after the initial detection of the vulnerability \cite{thung2012would}.

\begin{figure}[!htbp]
    \centering
    \includegraphics[width=1.0\linewidth]{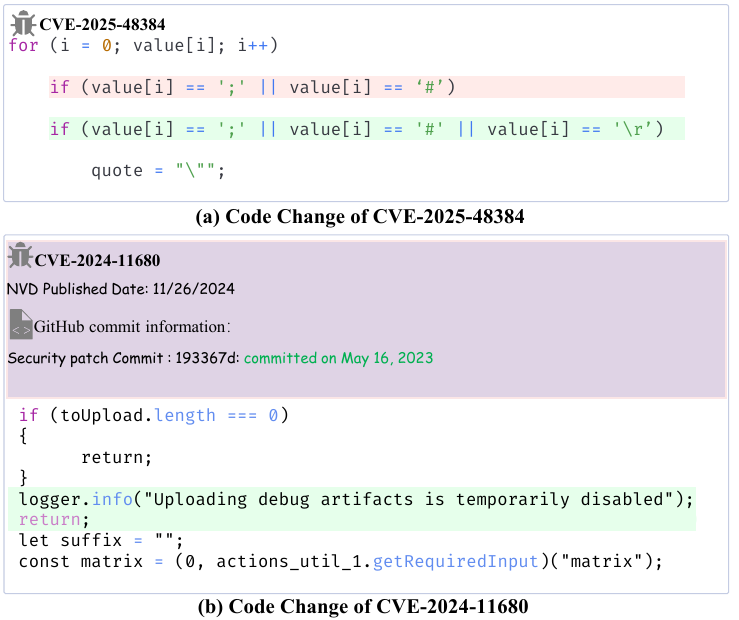}
    \caption{The part of the security patch that fixes CVE-2025-48384 (a) and CVE-2024-11680 (b). The red and green lines represent the before-fixed code (pre-patch) and after-fixed code (post-patch), respectively.}
    \label{fig:example}
\end{figure}

For instance, the CVE-2025-48384 vulnerability, as depicted in Figure \ref{fig:example} (a), was initially disclosed on July 8th. However, attackers may exploit it prior to its official disclosure. By injecting malicious files, including compromised Git Hook scripts, into local code repositories, attackers introduced severe security risks. Furthermore, Figure \ref{fig:example} (b) highlights the CVE-2024-11680 vulnerability, which enabled attackers to execute unauthorized operations. Although a patch for this vulnerability was already released in May 2023, most users failed to apply it promptly due to its silent patch releases. Consequently, when the CVE was officially published in November 2024, most affected software systems remained vulnerable. During this period, attackers widely exploited the vulnerability to modify server configurations and deploy WebShells, establishing remote control, ultimately causing multiple security incidents. Therefore, it is imperative for users and developers to adopt mechanisms capable of automatically and promptly distinguishing security patches from other updates, thereby mitigating security risks.

Deep learning (DL)-based approaches have demonstrated significant success in the detection of security patches, as these methods are capable of capturing complex dependency information and generalizing effectively across various types of vulnerabilities \cite{cheng2022bug,zhang2023multi,liu2024survey}. For instance, PatchRNN \cite{wang2021patchrnn} takes commit messages and code changes as input and employs a Recurrent Neural Network (RNN) \cite{elman1990finding} to process the input as sequences. Similarly, GraphSPD \cite{wang2023graphspd} and RepoSPD \cite{wen2024repository} represent code changes as graph structures that encode control-flow \cite{huo2020control} and data-flow \cite{cummins2021programl} information. They leveraged Graph Neural Networks (GNNs) \cite{li2015gated} or graph serialization techniques to identify security patches effectively. However, DL-based methods may still struggle to fully leverage the rich contextual information embedded in the input data and face challenges in learning patch representation due to the complex relationships among multiple code changes.

In recent years, Large Language Models (LLMs) have achieved remarkable success across a wide range of domains \cite{huang2024can}, including code generation \cite{zhang2025llm,pan2025codecor,wang2025teaching}, code repair \cite{li2025context,huang2025comprehensive,kong2025demystifying}, and vulnerability detection \cite{hussain2025vulbinllm,ding2025smartguard,tihanyi2025vulnerability}. Simultaneously, LLM-based agents have demonstrated substantial potential in software engineering \cite{he2025llm}, making notable advancements in areas such as requirements engineering \cite{ataei2025elicitron,jin2024mare}, code documentation generation \cite{yang2025docagent,luo2024repoagent}, and software testing \cite{yuan2024evaluating, deng2024pentestgpt}. By leveraging extensive pre-training on natural language corpora and large-scale code \cite{brown2020language,roziere2023llamacode}, along with the integration of advanced agent frameworks, LLMs and LLM-based agents have developed a robust understanding of natural languages and programming structures. This capability enables them to perform effectively in software security tasks such as vulnerability detection \cite{yu2025preliminary,wang2025vulagent}. However, systematic research on the evaluation of LLMs and LLM-based agents for security patch detection remains limited.

To bridge this gap, we conduct a comprehensive evaluation to investigate the effectiveness of LLMs and LLM-based agents in security patch detection. Specifically, we first study the effectiveness of three methods: Plain LLM, Data-Aug LLM, and React Agent. The Plain LLM employs a single LLM with a system prompt, while the Data-Aug LLM enhances the Plain LLM by incorporating external information through data augmentation. The React Agent, on the other hand, integrates thought-action-observation mechanisms to detect security patches. Building on this foundation, we evaluate the performance of various LLMs within these methods, including commercial LLMs of GPT-4o, GPT-4o-mini, GPT-5, DeepSeek-R1, and open-source LLMs like Llama-3.1 and Gemma-3. To further verify the effectiveness of LLMs in this task, we perform a comprehensive comparison of the studied methods against existing baselines. Additionally, we assess the performance of our studied methods across
different types of security vulnerabilities. We also explore the impact of various prompting strategies on model performance, including Chain-of-Thought (CoT), Few-Shot (FS), and the combination of CoT and FS (FS+CoT). Finally, we analyze how context window sizes influence detection accuracy, providing insights into the role of key parameters in determining overall performance.

Based on our experimental results, we generate the following key findings: 1) Data-Aug LLM demonstrates superior overall performance in the security patch detection task, whereas ReAct Agent achieves the best false positive rate (FPR) due to its thought, action, and observation process. 2) Commercial LLMs, such as GPT-4o, GPT-4o-mini, and DeepSeek-R1, consistently outperform open-source LLMs like Llama-3.1 and Gemma-3 across Plain LLM, Data-Aug LLM, and ReAct Agent methods. 3) Although our baselines exhibit strong performance in terms of accuracy and F1 score for the task of security patch detection, they demonstrate notable limitations in precision and FPR, where the GPT-4o model under the ReAct-Agent method achieves a high precision of 86.15\% and significantly reduces the FPR to 14.4\%. 4) All methods effectively identify security patches across various Common Weakness Enumeration (CWE) types, and all evaluated methods achieve their highest precision in the CWE-20 (Input Validation) category, their highest accuracy and F1 score in the CWE-264 (Authorization Management) category. 5) In most cases, the CoT prompting strategy is the most effective for improving model performance, followed by the hybrid strategy combining CoT and FS. 6) Models with larger context window sizes generally demonstrate enhanced performance across various metrics, as evidenced by LLaMA-3.1 and Gemma-3.

In summary, this paper makes the following contributions:

\begin{itemize}
    \item To the best of our knowledge, this study is the first systematic evaluation of LLMs and LLM-based agents in the task of security patch detection. By addressing a critical gap in the field, our work provides valuable insights and practical guidance for future research.
    \item We conduct a comprehensive investigation into the performance of the evaluated methods across various vulnerability types. Furthermore, we examine the impact of different prompting strategies and context window sizes on the effectiveness of security patch detection.
    \item To facilitate further research and advancements in this field, we have made our dataset and code publicly accessible at https://github.com/fzqn/PatchDetection.
\end{itemize}

The organization of this paper is as follows. Section \ref{related} describes the related work. Section \ref{Methodology} introduces the experimental setup. Section \ref{result} reports the experimental results. Section \ref{discuss} discusses implications for researchers and practitioners, and Section \ref{threats} presents threats to validity. Finally, we conclude the paper and discuss the future work in Section \ref{con}.

\section{Related Work}\label{related}
\subsection{Security patch detection}

Security patch detection is critical for enabling users to identify and apply updates that address vulnerabilities on time \cite{vaniea2016tales}. Initial approaches primarily relied on rule-based heuristic methods \cite{huang2019using, wu2020precisely} and traditional machine learning techniques \cite{tian2012identifying, xu2020automatic, wang2020machine, soto2016deeper, corley2011recovering}. For instance, Li et al. \cite{li2017large} conducted an empirical study on security patches, unveiling several key characteristics. Wu and Huang et al. \cite{wu2020precisely} developed rule-based approaches to identify common patterns in security patches. Moreover, Wang et al. \cite{wang2020machine} employed a random forest algorithm combined with extracted patch features to determine the vulnerability type associated with a specific security patch.

Subsequent research has adopted deep learning-based approaches for security patch detection \cite{zhou2021spi, nguyen2022vulcurator, wu2022enhancing, zhou2023colefunda}. Among these, Zuo et al. \cite{zuo2023commit} proposed a transformer-based detection approach that highlights the importance of commit messages in identifying security patches. Wang et al. \cite{wang2021patchrnn} introduced PatchRNN, a model that enhances patch identification performance by jointly modeling source code and commit messages. Zhou et al. \cite{zhou2021finding} proposed VulFixMiner, which extracts added and deleted code changes from commit messages and utilizes the CodeBERT model to identify security patches in Java and Python projects. Wang et al. \cite{wang2023graphspd} presented GraphSPD, a method that incorporates control flow graphs (PatchCPG) and leverages graph neural networks to improve the performance of security patch detection. Furthermore, Wen et al. \cite{wen2024repository} proposed RepoSPD, a repository-level security patch detection framework. This framework constructs a repository-level graph module, RepoCPG, and integrates structure-aware patch representations with a progressive learning mechanism to enable effective repository-level security patch detection. Zhou et al. \cite{zhou2023colefunda} proposed CoLeFunDa, a framework composed of a contrastive learner and FunDa, where FunDa is a novel function-change data augmentation method, to identify silent vulnerability fixes.

\subsection{LLM-based patch analysis}

Recent studies have explored the use of LLMs for patch detection, highlighting their ability to enhance the identification of vulnerability patches. Specifically, Tang et al. \cite{tang2023just} introduced LLMDA, using LLMs to generate code change explanations, and combining LLM outputs with representation learning techniques to enhance the identification of repair patches for security vulnerabilities. Li et al. \cite{li2024patchfinder} proposed PatchFinder, a two-stage framework incorporating initial retrieval and re-ranking, which effectively leverages the strengths of information retrieval (IR) and LLMs to achieve more accurate tracking of security patches. Luo et al. \cite{luo2024strengthening} developed SPatch, which employs fine-grained patch analysis and a novel differential symbolic execution technique to detect fine-grained security patches. Tian et al. \cite{tian2023best} investigated various code change representation learning methods to derive embeddings suitable for similarity computation in patch correctness identification. They further evaluated the potential of combining learned embeddings with engineered features for accurately classifying correct patches. Li et al. \cite{li2025they} developed DUALLM, a dual-method pipeline that integrates LLMs with fine-tuned small language models to enhance the performance of fine-grained patch classification. 

Simultaneously, except for the patch detection, researchers have also explored the characteristics of patches and the applications of LLMs in automated patch generation. For instance, Xie et al. \cite{xie2024unveiling} conducted a systematic study of the evolution of security patches in open-source projects, revealing the frequency and patterns of patch evolution and their impact on downstream tools for 1-day vulnerability analysis. Li et al. \cite{li2025empirical} constructed a large-scale binary patch dataset and systematically evaluated the capabilities of 19 code language models of varying scales on the task of binary security patch detection. Wang et al. \cite{wang2023rap} introduced RAP-Gen, a novel retrieval-augmented patch generation framework. This framework retrieves relevant repair patterns from a historical defect-repair codebase to enhance the generation capabilities of the CodeT5 patch generator. Lin et al. \cite{lin2024one} proposed Mulpor, a multi-granularity patch generation method that produces patches at the statement, expression, and token levels to address diverse real-world bugs.

However, existing studies have leveraged LLMs for information enhancement, employed LLMs for patch identification, or utilized them for patch generation. The application of LLMs in the domain of security patch detection remains unexplored, let alone the potential of LLM-based agents. Therefore, our study focuses on investigating the capabilities and performance of LLMs and LLM-based agents in the task of security patch detection.

\section{Experimental Setup} \label{Methodology}
\subsection{Overview}\label{Overview}

\begin{figure*}[t]
    \centering
    \includegraphics[width=0.9\textwidth]{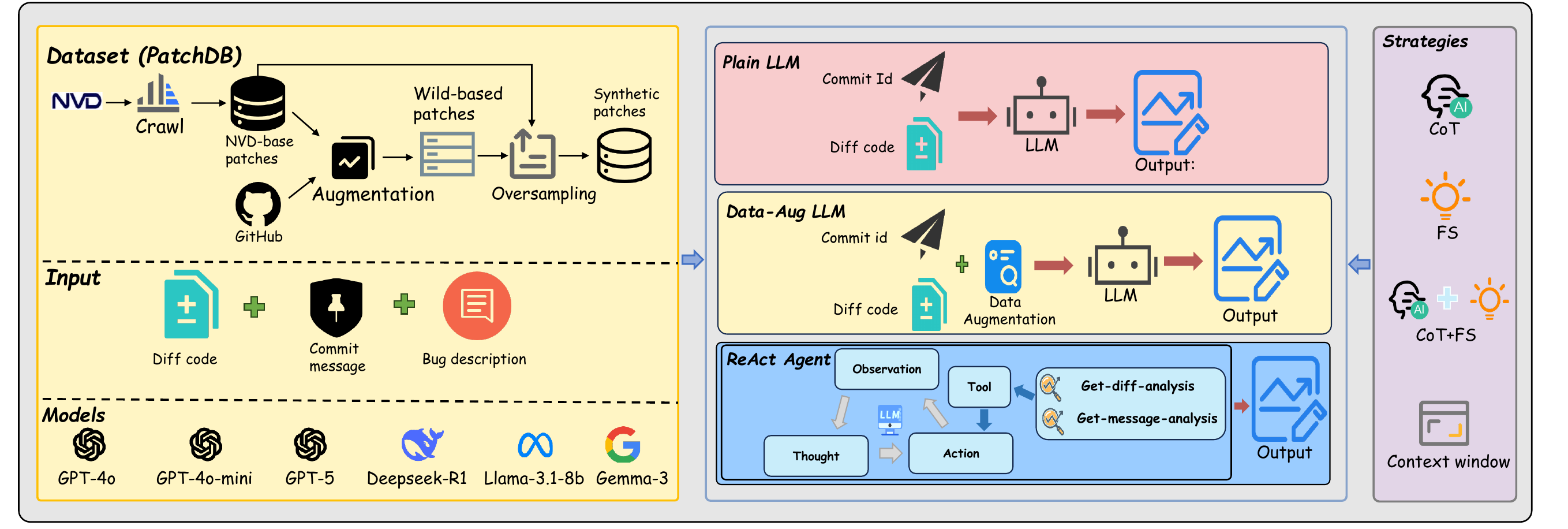}
    \caption{Overview of our approach.}
    \label{overview}
\end{figure*}

Figure~\ref{overview} presents a comprehensive overview of our experimental pipeline. The process begins with data collection, followed by the introduction of the studied methods, which include Plain LLM, Data-Augmented LLM, and React Agent. Subsequently, a diverse set of LLMs is selected for evaluation. Using these models, we systematically evaluate the studied methods by comparing them against baselines, examining their effectiveness across different CWE types, and analyzing the impact of various prompting strategies, such as Chain-of-Thought (CoT), Few-Shot (FS), and a hybrid approach combining CoT and FS (FS+CoT). Additionally, we investigate the influence of context window sizes on performance. Finally, the evaluation metrics employed in this study are presented.

\subsection{Research Questions} \label{RQ}

In this paper, we aim to answer the following research questions (RQ):

\begin{itemize}
    \item \textbf{RQ1:} How do different methods perform in security patch detection?
    \item \textbf{RQ2:} How do different LLMs perform in security patch detection?
    \item \textbf{RQ3:} How effective are LLMs and LLM-based agents compared with existing approaches in security patch detection?
    \item \textbf{RQ4:} How effective are LLMs and LLM-based agents in performing over patches with different vulnerability types?
    \item \textbf{RQ5:} How do different prompting strategies influence the effectiveness of LLMs and LLM-based agents?
    \item \textbf{RQ6:} Does the context window (CW) size affect the overall effectiveness of LLMs and LLM-based agents?
\end{itemize}

\subsection{Data Collection} \label{dataset}

To address the proposed RQs, our study draws inspiration from prior work \cite{wang2023graphspd,wen2024repository} and utilizes the widely adopted dataset of PatchDB \cite{wang2021patchdb}. PatchDB aggregates data from 313 open-source repositories, encompassing 12,000 real-world security patches, and 23,000 rigorously curated non-security patches. The dataset comprises three components, namely, the NVD-based dataset extracted from reference hyperlinks in the National Vulnerability Database (NVD), the wild-based dataset collected from GitHub commit histories, and the synthetic dataset that employs an innovative oversampling technique to synthesize patches at the source code level by enriching the control flow variants of the original patches. This comprehensive structure ensures robust coverage and diversity for patch-related research.

This dataset includes records with various attributes, including ``category", ``source", ``CVE-ID", ``CWE-ID", ``owner", ``repo", ``commit-id", ``commit-message", and ``diff-code". Specifically, the ``category" attribute represents the type of patch, indicating whether it is a security patch or a non-security patch. The ``source" attribute specifies the origin of the patch, denoting whether the record has been extracted from the NVD, the wild, or synthetic sources. The ``CVE-ID" attribute corresponds to the Common Vulnerabilities and Exposures (CVE) identifier, formatted as ``CVE-XXXX-XXXXX" if available, or marked as ``NA" otherwise. Similarly, the ``CWE-ID" field refers to the CWE identifier, which is either a valid CWE ID or ``NA" in cases where it is not applicable. The ``owner" attribute identifies the owner of the repository, while ``repo" specifies the repository's unique identifier to which the record belongs. The ``commit-id" is the hash value of the corresponding commit. The ``commit-message" provides a description of the patch, and the ``diff-code" captures the diff code of the patch.

\subsection{Experimental Pipeline} \label{set}

In this section, we present our experimental pipeline, which encompasses the methodologies and configurations to address each RQ. Specifically, we outline the studied methods to address RQ1, the model selection and configurations to answer RQ2, the baselines employed for RQ3, the distribution of CWE Types for RQ4, the prompting strategies utilized to tackle RQ5, and the context window configurations implemented to address RQ6.

\subsubsection{Studied Methods}\label{model}

Our study explores three categories of detection methods: Plain LLM, Data-Aug LLM, and ReAct Agent. We chose the ReAct Agent due to its thought-action-observation workflow, which is aligned with the need for handling complex tasks such as security patch detection. This approach enables more effective reasoning and decision-making, making it highly compatible with the intricate demands of identifying and analyzing patches across diverse scenarios.

\begin{figure}[!t]
    \centering
    \includegraphics[width=3.5in]{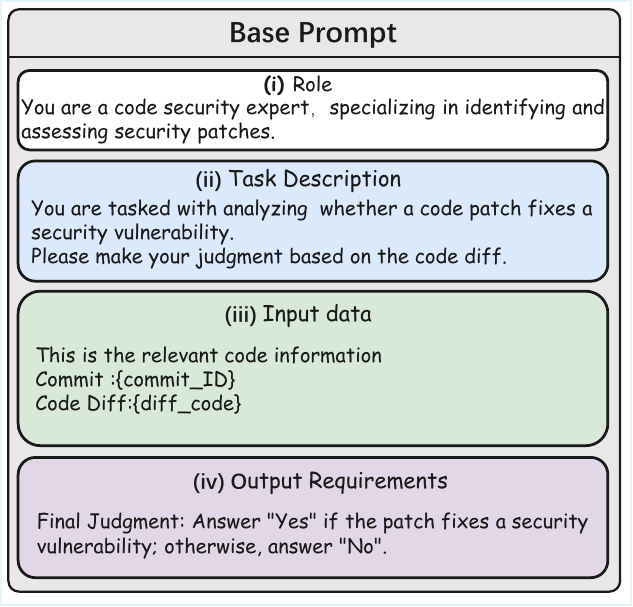}
    \caption{The prompt template used with Plain LLM.}
    \label{fig:sysprompt}
\end{figure}

\textbf{Plain LLM} approach utilizes a single LLM with a system prompt to detect security patches. Specifically, the LLM processes the code differences (code diffs) in a commit as input and determines whether the commit is a security patch. The detailed design of the prompt is illustrated in Figure \ref{fig:sysprompt}.

\begin{figure}[h]
    \centering
    \includegraphics[width=3.5in]{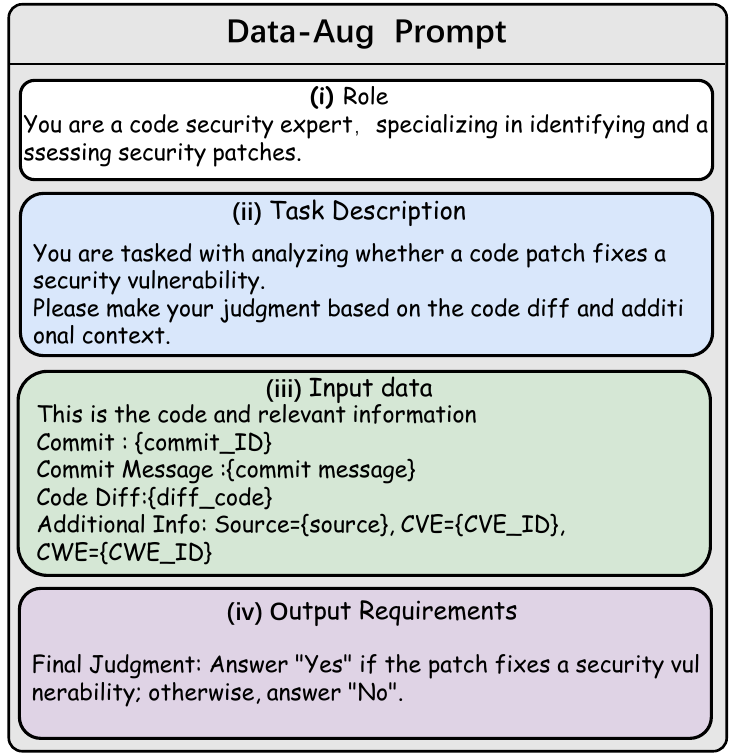}
    \caption{The prompt template used with Data-Aug LLM.}
    \label{fig:dataprompt}
\end{figure}

\textbf{Data-Aug LLM} builds upon the Plain LLM by augmenting the code diffs in each commit with additional commit-related information to enhance the detection performance of security patches. Specifically, the enriched information includes the commit's detailed description, its source (e.g., GitHub or the NVD website), the CVE identifier associated with the code, and the corresponding vulnerability's CWE category. This approach aims to evaluate whether LLMs equipped with enriched contextual information can effectively identify security patches. The detailed prompt design is presented in Figure \ref{fig:dataprompt}.

\begin{figure}[h]
    \centering
    \includegraphics[width=3.5in]{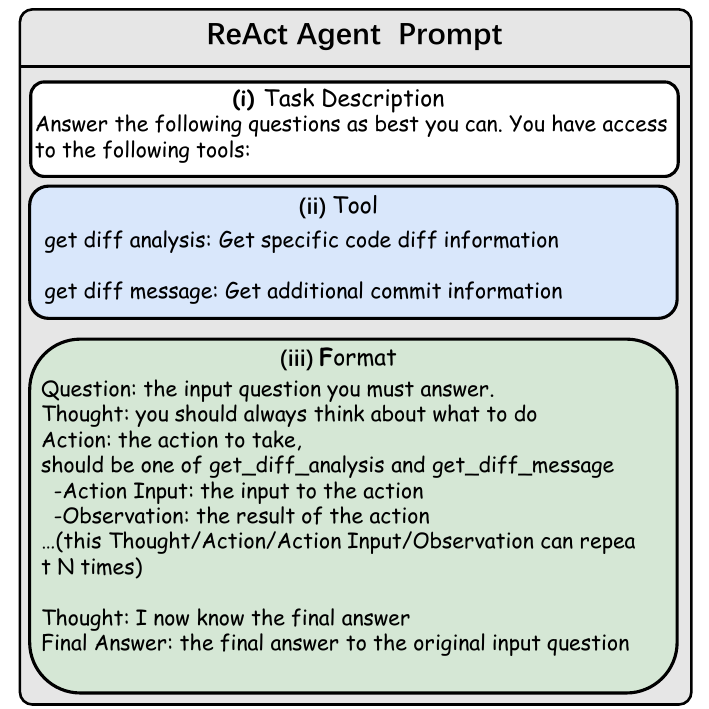}
    \caption{The prompt template used with ReAct Agent.}
    \label{fig:agentprompt}
\end{figure}

\textbf{ReAct Agent} employs an iterative thought-action-observation framework to detect security patches. It integrates two essential tools for on-demand interprocedural context acquisition: (i) get\_diff\_analysis, which provides detailed insights into code differences, and (ii) get\_diff\_message, which extracts commit-related information. Specifically, the ReAct agent first receives the original code and the modified code of a commit as input. During each iteration, the agent reasons based on the original code and the modified code, as well as observations from prior iterations. It then determines whether to invoke the tools for interprocedural context acquisition or to terminate the iteration and make a final prediction. After observing the tool outputs, the agent proceeds to the next iteration. Figure \ref{fig:agentprompt} presents the detailed prompt design.

\subsubsection{Model Selection and Configurations}\label{model}

\noindent\textbf{Model Selection:} As shown in Table \ref{tab:modelselect}, we evaluate the performance of various LLMs across three studied methods, i.e., the Plain LLM, Data-Aug LLM, and React Agent, on the security patch detection task. The evaluated models are categorized into two groups: commercial LLMs and open-source LLMs. For commercial LLMs, we examine GPT-4o, GPT-4o-mini, GPT-5, and DeepSeek-R1, while for open-source LLMs, we include Llama-3.1 and Gemma-3. Specifically, Llama-3.1-8B is used as the representative version of Llama-3.1, while Gemma-3-7B serves as the implementation version for Gemma-3. The rationale we selecting these models is based on their remarkable performance across various tasks and their timeliness.

\begin{table}[htbp]
  \centering
  \caption{LLMs selection.}
    \begin{tabular}{p{7.19em}p{4.19em}cp{4.19em}p{4.19em}}
    \toprule
    \textbf{Model} & \textbf{Size} & \multicolumn{1}{p{4.19em}}{\textbf{Time}} & \textbf{Instruct} & \textbf{Base} \\
    \midrule
    GPT-4o & N/A   & 2024/5/13 & \textbf{×} & \checkmark \\
    \rowcolor[rgb]{ .949,  .949,  .949} GPT-4o-mini & N/A   & 2024/7/28 & \textbf{×} & \checkmark \\
    GPT-5 & N/A  & 2025/8/7 & \textbf{×} & \checkmark \\
    \rowcolor[rgb]{ .949,  .949,  .949} Deepseek-R1 & 671K  & 2025/1/20 & \textbf{×} & \checkmark \\
    Llama-3.1 & 8B    & 2024/7/23 & \checkmark     & \textbf{×} \\
    \rowcolor[rgb]{ .949,  .949,  .949} Gemma-3 & 7B    & 2024/2/21 & \textbf{×} & \checkmark \\
    \bottomrule
    \end{tabular}%
  \label{tab:modelselect}%
\end{table}%

\noindent\textbf{Model Configurations:} To ensure reproducibility, consistency, and a fair comparison across the selected LLMs, our pipeline adopts carefully designed model configurations. Following the best practices outlined in Yang et al's study \cite{yang2025context}, we set the temperature to 0.1 for all open-source models. This reduces randomness and enables the generation of deterministic outputs, which is critical for security patch detection tasks that need precise answers while allowing the model to explain its reasoning. To enhance reproducibility, we fixed the random seed to 42 for all open-source models, as recommended by Lin et al.'s study \cite{lin2025large}. This ensures that, given the same pre-trained weights and input, the models produce consistent outputs across different runs. Additionally, we limited the maximum output length to 512 tokens for all open-source models to prevent excessive generation and mitigate potential memory issues. 

For commercial models, we retained their default configurations while aligning key parameters with those of the open-source models to maintain consistency. Specifically, the temperature was set to 0.1, and the maximum output length was set to 512 tokens.

\subsubsection{Baselines} \label{baseline}

To address RQ3, we compare our evaluated methods with several state-of-the-art approaches for security patch detection, including PatchFinder \cite{li2024patchfinder}, VCMatch \cite{wang2022vcmatch}, and RepoSPD \cite{wen2024repository}. PatchFinder \cite{li2024patchfinder} employed a two-stage detection framework. In the retrieval phase, it utilized a hybrid patch retriever to identify candidate commits that exhibit both lexical and semantic similarity to CVE descriptions. In the re-ranking phase, a fine-tuned model is applied to re-rank these candidates, enhancing detection accuracy. VCMatch \cite{wang2022vcmatch} extracted both statistical and semantic features of vulnerabilities and the patch commits, leveraging a voting-based ranking fusion method to combine the outputs of three classification models-XGBoost, LightGBM, and CNN-for optimal results. RepoSPD \cite{wen2024repository} introduced a repository-level security patch detection framework comprising three key components: repository-level graph construction, structure-aware patch representation, and progressive learning to enhance security patch detection.

As this study focuses on a binary classification task, we made adjustments to the outputs of the baselines to ensure a fair comparison. PatchFinder retains its two-stage architecture, but the re-ranking stage was modified to produce binary outputs. Similarly, VCMatch was adapted by adjusting its classification layer to generate binary outputs. In contrast, RepoSPD inherently supports end-to-end binary classification and requires no structural modifications.

\subsubsection{Distribution of CWE Types}\label{type}

\begin{table}[!ht]
  \centering
  \caption{CWE types and corresponding quantities.}
  \renewcommand{\arraystretch}{1.15}
  \setlength{\tabcolsep}{6pt}
  \scriptsize
  \begin{tabular}{
    >{\raggedright\arraybackslash}p{5em} 
    >{\raggedright\arraybackslash}p{13em} 
    >{\raggedleft\arraybackslash}p{5em}}
    \toprule
    \textbf{CWE-ID} & \textbf{CWE Name} & \textbf{Quantity} \\
    \midrule
    \rowcolor[HTML]{F7F7F7}119 & Information Exposure  & 926 \\
    20 & Input Validation & 432 \\
    \rowcolor[HTML]{F7F7F7}399 & Resource Management Errors & 288 \\
    200 & Information Exposure & 267 \\
    \rowcolor[HTML]{F7F7F7} 125 & Out-of-Bounds Read & 261 \\
    264 & Authorization Management & 240 \\
    \rowcolor[HTML]{F7F7F7}189 & Numeric Errors & 234 \\
    476 & NULL Pointer Dereference & 161 \\
    \rowcolor[HTML]{F7F7F7} 190 & Integer Overflow  & 133 \\
    416 & Use After Free & 103 \\
    \rowcolor[HTML]{F7F7F7} Other & Other CWE Types & 708 \\
    \bottomrule
  \end{tabular}
  \label{tab:CWEtypes}
\end{table}

To answer RQ4, we evaluate the effectiveness of our methods in identifying various types of security patches by utilizing the same dataset employed in the previous RQs, which includes the information of CWE types. To ensure a focused analysis, we retain the top 10 CWE types with the highest number of samples, and group all remaining CWE types into an ``Other" category. The distribution of these CWE types is summarized in Table \ref{tab:CWEtypes}.

\subsubsection{Prompting Strategies}

For each detection method, in addition to the standard version utilizing a basic prompt, we design three variants based on different prompting strategies: Chain-of-Thought (CoT), Few-Shot Learning (FS), and a hybrid approach combining both CoT and FS (CoT + FS). Notably, we use \textbf{Vanilla} to denote the absence of any prompting strategy.

\begin{figure}[!h]
    \centering
    \includegraphics[width=3.5in]{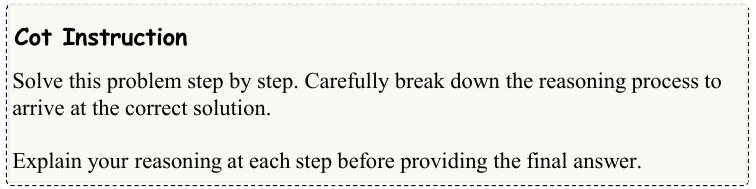}
    \caption{The prompting strategies of CoT.}
    \label{fig:cotprompt}
\end{figure}

\textbf{CoT}: A CoT prompting strategy is adopted by incorporating the instruction ``solve the problem step-by-step and provide corresponding reasoning" into the prompt, where the prompt design is illustrated in Figure \ref{fig:cotprompt}. This strategy guides the LLM and ReAct agent to systematically break down reasoning tasks. Following insights from prior work \cite{yildiz2025benchmarking}, we intentionally avoid more complex CoT instructions, as summarizing reasoning patterns for vulnerabilities and security patches in advance proves to be challenging, particularly given the extensive diversity of vulnerability types (CWEs).

\begin{figure}[!h]
    \centering
    \includegraphics[width=3.5in]{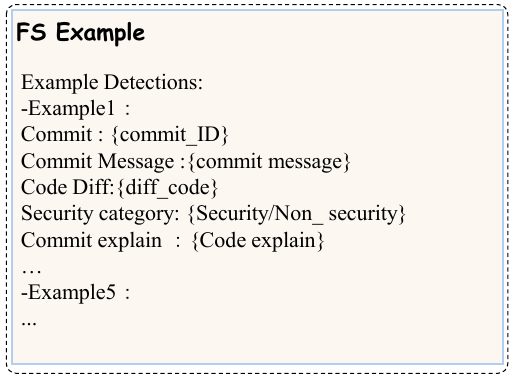}
    \caption{The prompting strategies of FS.}
    \label{fig:FSprompt}
\end{figure}

\textbf{FS}: The prompt includes several annotated examples, each comprising a vulnerable code snippet, the corresponding commit message, a code diff, an annotation indicating whether this is a security patch, and the rationale behind the annotation. These examples are expected to guide the LLM and ReAct agent in accurately identifying security patches while emphasizing the defining characteristics of security patches. The prompt design is presented in Figure \ref{fig:FSprompt}.

\textbf{CoT + FS}: A hybrid approach that combines CoT reasoning with FS examples, aiming to leverage the strengths of both strategies.

\subsubsection{Context Window Configurations}

The context window (CW) size represents the maximum number of tokens a LLM can process in a single inference. Prior research \cite{lin2025large} shows that the size of the CW directly affects LLMs' ability to capture long dependencies, incorporate contextual information, and maintain coherent reasoning. In the security patch detection task, inputs often consist of complex code changes and detailed descriptions that exceed input limits, making CW size a critical determinant of model performance. Smaller CWs risk truncating essential information and thereby reducing detection accuracy, whereas larger CWs enable models to capture more extensive context but may introduce noise and incur increased computational overhead. 

To systematically examine this trade-off, we configure CW sizes as follows: for closed-source models, DeepSeek-R1 supports up to 64K tokens, GPT-4o and GPT-4o-mini accommodate 128K tokens, and GPT-5 supports 40K tokens; for open-source models, Llama-3.1 and Gemma-3 support 2,048, 4,096, and 8,192 tokens, respectively. These settings balance the need for evaluating the impact of CW size on detection performance and addressing practical constraints, such as computational resources and model compatibility. By systematically examining CW size, this study aims to provide deeper insights into its role in optimizing model performance.

\subsection{Evaluation Metrics} \label{metrics}

The task of security patch detection is fundamentally a binary classification problem. To comprehensively evaluate the performance of studied methods on security patch detection, we employ four widely adopted evaluation metrics in prior studies \cite{wang2021patchrnn} \cite{wen2024repository} \cite{fu2022linevul} \cite{wang2023graphspd}: precision, accuracy, F1 Score, and false positive rate (FPR).

\section{Evaluation Results} \label{result}
\subsection{RQ1: How do different methods perform in security patch detection?}

Table \ref{table:result} presents a comparative analysis of three research methods—Plain LLM, Data-Aug LLM, and ReAct Agent—in the task of security patch detection. The results reveal that Data-Aug LLM achieves the highest average precision of 81.67\%, surpassing ReAct Agent with 79.53\% and Plain LLM with 77.98\%, representing improvements of 2.14\% and 3.69\%, respectively. Additionally, Data-Aug LLM demonstrates superior performance in terms of average accuracy (57.48\%) and average F1 score (56.14\%), outperforming Plain LLM's 54.16\% and 52.30\% as well as ReAct Agent's 51.98\% and 47.14\%. This improvement may be attributed to its context augmentation mechanism, which incorporates supplementary information, such as commit messages, to enhance the model's ability to identify true positives. However, this mechanism may also introduce noise, resulting in an increased false positive rate of 23.96\%, which is notably higher than ReAct Agent's 14.14\% but slightly above Plain LLM's 22.56\%. 

The advantage of ReAct Agent, on the other hand, may result from its iterative thought-action-observation mechanism, which progressively filters out non-security patches and enhances decision reliability. In contrast, Plain LLM lacks both contextual augmentation and task-specific reasoning capabilities, resulting in its comparatively lower performance across all evaluated metrics.  

\begin{table}[!htbp]
    \centering
    \caption{Performance comparison of three research methods—Plain LLM, Data-Aug LLM, and ReAct Agent—in the task of security patch detection. The bold value indicates the optimal performance. PRE refers to precision, ACC refers to accuracy, F1 represents the F1 score, and FPR denotes the false positive rate.}
    \includegraphics[width=1.0\linewidth]{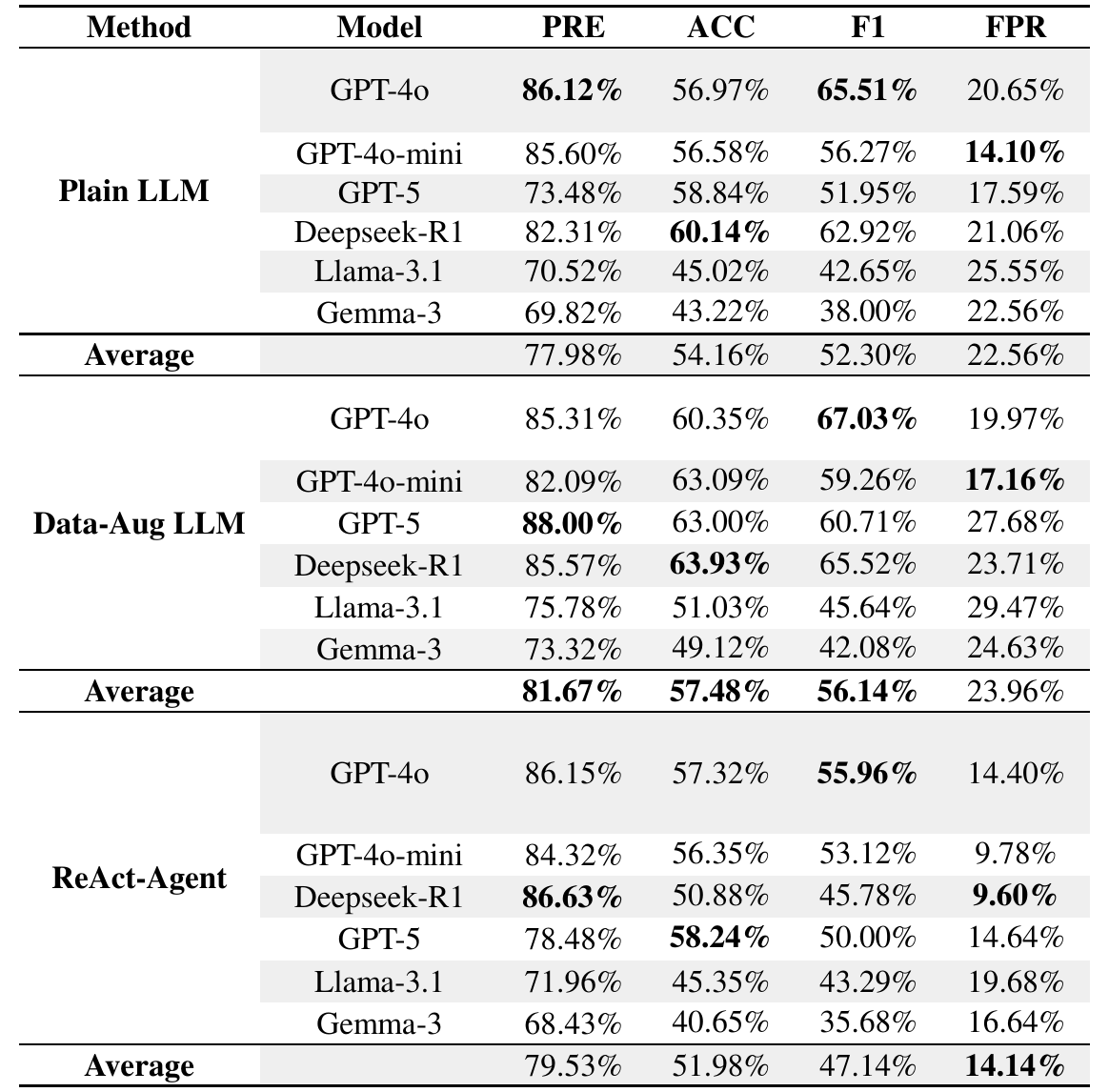}
    \label{table:result}
\end{table}

\begin{tcolorbox}[answerbox]
  \textbf{Finding 1:} Leveraging its data augmentation mechanism, Data-Aug LLM demonstrates superior overall performance in the security patch detection task, whereas ReAct Agent achieves the lowest false positive rate (FPR) due to its iterative thought-action-observation process. 
\end{tcolorbox}

\subsection{RQ2: How do different LLMs perform in security patch detection?}

Table \ref{table:result} reveals significant differences among the evaluated LLMs. Within the method of Plain LLM, GPT-4o achieves the highest precision and F1 score, reaching 86.12\% and 65.51\%, respectively, followed by GPT-4o-mini and DeepSeek-R1. In terms of accuracy, DeepSeek-R1 leads with 60.14\%, closely followed by GPT-5 at 58.84\%. Notably, GPT-4o-mini achieves the lowest false positive rate at 14.10\%, with GPT-5 ranking second. Overall, commercial LLMs such as GPT-4o, GPT-4o-mini, and DeepSeek-R1 demonstrate superior performance. In contrast, open-source LLMs like Llama-3.1 and Gemma-3 exhibit comparatively lower performance, likely due to their smaller model size or limited domain-specific adaptability.

Under the Data-Aug LLM method, GPT-5 achieves the highest precision at 88.00\%, followed by DeepSeek-R1. In terms of accuracy, DeepSeek-R1 leads with 63.93\%, with GPT-4o-mini as the second. Meanwhile, GPT-4o achieves the highest F1 score at 67.03\%, followed by DeepSeek-R1. Additionally, GPT-4o-mini demonstrates the lowest false positive rate at 17.16\%, with GPT-4o ranking second. Overall, commercial LLMs, including GPT-5, GPT-4o, GPT-4o-mini, and DeepSeek-R1, consistently exhibit superior performance under the Data-Aug LLM method. In contrast, open-source LLMs such as Llama-3.1 and Gemma-3 perform comparatively poorly. Gemma-3 records the lowest precision, accuracy, and F1 score at 73.32\%, 49.12\%, 42.08\%, respectively, while Llama-3.1 shows the highest FPR at 29.47\%.

Furthermore, compared to the Plain LLM, the incorporation of data augmentation results in improved precision for all LLMs except GPT-4o and GPT-4o-mini, with GPT-5 exhibiting the most significant improvement at 14.52\%. Additionally, all models demonstrate notable improvements in both accuracy and F1 score, with accuracy improvements ranging from 3.38\% to 6.51\% and F1 score improvements ranging from 1.52\% to 8.76\%. These results highlight the importance of data augmentation in enhancing performance for the security patch detection task. However, the introduction of additional information may also introduce noise, leading to increased FPR for all LLMs except GPT-4o when compared to the Plain LLM. Future research may focus on balancing accuracy and FPR to achieve optimal performance.

Under the ReAct Agent method, DeepSeek-R1 demonstrates exceptional performance, achieving the highest precision at 86.63\% while maintaining the lowest false positive rate at 9.60\%. This indicates that under this method, DeepSeek-R1 demonstrates optimal capabilities and achieves a balance between accuracy and false positive rate. Meanwhile, GPT-5 achieves the highest accuracy at 58.24\%, and GPT-4o attains the highest F1 score at 55.96\%. In contrast, open-source LLMs continue to display comparatively weaker performance. 

Moreover, compared with the Plain LLM and Data-Aug LLM methods, although most LLMs under the Data-Aug LLM method achieve the highest precision, accuracy, and F1 score, LLMs under the ReAct Agent method demonstrate the lowest false positive rate. This highlights the importance of integrating iterative reasoning strategies into the Data-Aug LLM method to further optimize performance while simultaneously achieving a balance between accuracy and false positive rate.

\begin{tcolorbox}[answerbox]
  \textbf{Finding 2:} The evaluation results reveal that commercial LLMs, such as GPT-4o, GPT-4o-mini, and DeepSeek-R1, consistently outperform open-source LLMs like Llama-3.1 and Gemma-3 across Plain LLM, Data-Aug LLM, and ReAct Agent methods. Meanwhile, although most LLMs under the Data-Aug method achieve the highest precision, accuracy, and F1 score, the LLMs under the ReAct Agent method achieve the lowest false positive rates. This highlights the potential of combining data augmentation with iterative reasoning strategies to optimize performance while achieving a balanced trade-off between accuracy and false positive rates.
\end{tcolorbox}

\subsection{RQ3: How effective are LLMs and LLM-based agents compared with existing approaches in security patch detection?}

\begin{table}[!htbp]
  \centering
  \caption{Comparison of the optimal and average performance of our evaluated methods with baselines. PRE refers to precision, ACC refers to accuracy, F1 represents the F1 score, and FPR denotes the false positive rate.}
  \renewcommand{\arraystretch}{1.05}
  \setlength{\tabcolsep}{3.5pt} 
  \resizebox{\columnwidth}{!}{ 
  \begin{tabular}{>{\centering\arraybackslash}p{2.9cm}
                  >{\centering\arraybackslash}p{1.3cm}
                  >{\centering\arraybackslash}p{1.3cm}
                  >{\centering\arraybackslash}p{1.3cm}
                  >{\centering\arraybackslash}p{1.3cm}}
    \toprule
    \textbf{Method} & \textbf{PRE (\%)} & \textbf{ACC (\%)} & \textbf{F1 (\%)} & \textbf{FPR (\%)} \\
    \midrule
    PatchFinder & 75.17 & \textbf{75.30} & \textbf{83.54} & 45.86 \\
    \rowcolor[gray]{0.96} VCMatch & 82.64 & 74.30 & 80.15 & 32.73 \\
    RepoSPD & 83.13 & 69.98 & 68.13 & 14.57 \\
    \rowcolor[gray]{0.96} \makecell[c]{Plain LLM\\(Optimal)} & 86.12 & 56.97 & 65.51 & 20.65 \\
    \makecell[c]{Plain LLM\\(Average)} & 77.98 & 54.16 & 52.30 & 22.56 \\
    \rowcolor[gray]{0.96} \makecell[c]{Data-Aug LLM\\(Optimal)} & 85.57 & 63.93 & 65.52 & 23.71 \\
    \makecell[c]{Data-Aug LLM\\(Average)} & 81.67 & 57.48 & 56.14 & 23.96 \\
    \rowcolor[gray]{0.96} \makecell[c]{ReAct-Agent\\(Optimal)} & \textbf{86.15} & 57.32 & 55.96 & 14.40 \\
    \makecell[c]{ReAct-Agent\\(Average)} & 79.53 & 51.98 & 47.14 & \textbf{14.14} \\
    \bottomrule
  \end{tabular}
  }
  \label{tab:baseline_result}
\end{table}

Table~\ref{tab:baseline_result} provides a comparative analysis of the best-performing models and the average performance of each method evaluated against existing baseline approaches. The results reveal that PatchFinder achieves the highest accuracy and F1 score, reaching 75.30\% and 83.54\%, respectively, demonstrating the effectiveness of feature extraction and representation in the task of security patch detection. Similarly, the Data-Aug LLM method, which leverages data augmentation, also demonstrates superior performance across these two metrics among the evaluated methods. These findings emphasize the importance of contextual information and feature representation in achieving effective security patch detection. 

However, PatchFinder exhibits a relatively high false positive rate, reaching 45.86\%, which suggests that relying solely on feature extraction and representation is insufficient to capture complex semantic changes, resulting in a substantial number of false positives. In contrast, among the methods we evaluated, the ReAct-Agent method achieves an average FPR of 14.14\%, with its best-performing model, GPT-4o, achieving an FPR of 14.4\%. These results further highlight the effectiveness of the iterative thought-action-observation mechanism employed by ReAct-Agent in security patch detection, particularly in significantly reducing false positive rates.

Furthermore, the GPT-4o model within the ReAct-Agent method achieves an accuracy of 86.15\%, which is 3\% to 11\% higher than existing baselines. This provides strong evidence of the effectiveness of the ReAct-Agent method, particularly the GPT-4o model, in the task of security patch detection. Notably, the RepoSPD method, which leverages code graphs to capture inter-code dependencies, performs comparably to LLM-based methods and slightly outperforms them in terms of accuracy and F1 score, while also maintaining relatively low precision and high FPR. This advantage likely arises from its ability to model structural relationships beyond textual representations. However, as its precision and FPR remain inferior to those of the ReAct-Agent method, future research could explore the integration of structural relationship modeling with iterative reasoning mechanisms to achieve superior performance while maintaining a low false positive rate.

\begin{tcolorbox}[answerbox]
  \textbf{Finding 3:} Although PatchFinder and RepoSPD exhibit strong performance in terms of accuracy and F1 score for the task of security patch detection, they demonstrate notable limitations in precision and false positive rates. In contrast, the results highlight the effectiveness of the ReAct-Agent method, particularly the GPT-4o model, in achieving a high precision of 86.15\% and significantly reducing the false positive rates to 14.4\%, outperforming other approaches.
\end{tcolorbox}

\subsection{RQ4: How effective are LLMs and LLM-based agents in performing over patches with different vulnerability types?}

\begin{table}[!htbp]
    \centering
    \caption{Performance of best-performing models across various CWE types.}
    \resizebox{1.0\linewidth}{!}{
        \begin{tabular}{c}
            \subfloat[Plain LLM]{

\begin{tabular}{cp{3.5em}cccc}
    \toprule
    \multicolumn{1}{p{2.1em}}{\textbf{Model}} & \multicolumn{1}{c}{\textbf{CWE-ID}} 
    & \multicolumn{1}{c}{\textbf{PRE (\%)}} & \multicolumn{1}{c}{\textbf{ACC (\%)}} 
    & \multicolumn{1}{c}{\textbf{F1 (\%)}} & \multicolumn{1}{c}{\textbf{FPR (\%)}} \\
    \midrule
    \multicolumn{1}{c}{\multirow{11}[2]{*}{\makecell{Plain LLM \\ (GPT-4o)}}} 
    & \multicolumn{1}{c}{119} & 90.50 & 86.38 & 79.95 & 9.67 \\
          & \multicolumn{1}{c}{20} & \cellcolor[rgb]{ .937,  .937,  .937}\textbf{92.32} & \cellcolor[rgb]{ .937,  .937,  .937}82.51 & \cellcolor[rgb]{ .937,  .937,  .937}77.61 & \cellcolor[rgb]{ .937,  .937,  .937}8.53 \\
          & \multicolumn{1}{c}{399} & 88.51 & 80.30 & 70.13 & 9.66 \\
          & \multicolumn{1}{c}{200} & \cellcolor[rgb]{ .937,  .937,  .937}78.93 & \cellcolor[rgb]{ .937,  .937,  .937}83.11 & \cellcolor[rgb]{ .937,  .937,  .937}77.59 & \cellcolor[rgb]{ .937,  .937,  .937}8.19 \\
          & \multicolumn{1}{c}{125} & 83.40 & 83.81 & 77.83 & 8.17 \\
          & \multicolumn{1}{c}{264} & \cellcolor[rgb]{ .937,  .937,  .937}89.81 & \cellcolor[rgb]{ .937,  .937,  .937}\textbf{91.37} & \cellcolor[rgb]{ .937,  .937,  .937}\textbf{86.88} & \cellcolor[rgb]{ .937,  .937,  .937}11.28 \\
          & \multicolumn{1}{c}{189} & 89.79 & 83.19 & 80.48 & 8.75 \\
          & \multicolumn{1}{c}{476} & \cellcolor[rgb]{ .937,  .937,  .937}81.49 & \cellcolor[rgb]{ .937,  .937,  .937}77.31 & \cellcolor[rgb]{ .937,  .937,  .937}74.12 & \cellcolor[rgb]{ .937,  .937,  .937}\textbf{8.02} \\
          & \multicolumn{1}{c}{190} & 86.74 & 87.49 & 79.89 & 10.92 \\
          & \multicolumn{1}{c}{416} & \cellcolor[rgb]{ .937,  .937,  .937}86.28 & \cellcolor[rgb]{ .937,  .937,  .937}79.34 & \cellcolor[rgb]{ .937,  .937,  .937}75.27 & \cellcolor[rgb]{ .937,  .937,  .937} 8.09 \\
          & \multicolumn{1}{c}{Other} & 89.00 & 77.64 & 79.60 & 9.72 \\
    \bottomrule
    \end{tabular}

            }\\[6pt]  
            \subfloat[Data-Aug LLM]{

\begin{tabular}{cp{3.0em}cccc}
    \toprule
    \multicolumn{1}{p{2.1em}}{\textbf{Model}} & \multicolumn{1}{c}{\textbf{CWE-ID}} 
    & \multicolumn{1}{c}{\textbf{PRE (\%)}} & \multicolumn{1}{c}{\textbf{ACC (\%)}} 
    & \multicolumn{1}{c}{\textbf{F1 (\%)}} & \multicolumn{1}{c}{\textbf{FPR (\%)}} \\
    \midrule
    \multicolumn{1}{c}{\multirow{11}[2]{*}{\makecell{Data-Aug LLM \\ (Deepseek-R1)}}} 
    & \multicolumn{1}{c}{119} & 89.17 & 84.32 & 77.34 & 10.02 \\
          & \multicolumn{1}{c}{20} & \cellcolor[rgb]{ .937,  .937,  .937}\textbf{90.45} & \cellcolor[rgb]{ .937,  .937,  .937}80.05 & \cellcolor[rgb]{ .937,  .937,  .937}74.76 & \cellcolor[rgb]{ .937,  .937,  .937}9.55 \\
          & \multicolumn{1}{c}{399} & 86.41 & 78.12 & 67.72 & 10.48 \\
          & \multicolumn{1}{c}{200} & \cellcolor[rgb]{ .937,  .937,  .937}76.81 & \cellcolor[rgb]{ .937,  .937,  .937}81.39 & \cellcolor[rgb]{ .937,  .937,  .937}75.50 & \cellcolor[rgb]{ .937,  .937,  .937} 9.14 \\
          & \multicolumn{1}{c}{125} & 81.05 & 81.57 & 75.63 & \textbf{8.75} \\
          & \multicolumn{1}{c}{264} & \cellcolor[rgb]{ .937,  .937,  .937}87.35 & \cellcolor[rgb]{ .937,  .937,  .937}\textbf{89.58} & \cellcolor[rgb]{ .937,  .937,  .937}\textbf{84.51} & \cellcolor[rgb]{ .937,  .937,  .937}12.74 \\
          & \multicolumn{1}{c}{189} & 88.01 & 81.88 & 78.77 & 9.57 \\
          & \multicolumn{1}{c}{476} & \cellcolor[rgb]{ .937,  .937,  .937}79.38 & \cellcolor[rgb]{ .937,  .937,  .937}75.09 & \cellcolor[rgb]{ .937,  .937,  .937}71.95 & \cellcolor[rgb]{ .937,  .937,  .937}9.38 \\
          & \multicolumn{1}{c}{190} & 84.30 & 85.49 & 77.69 & 11.35 \\
          & \multicolumn{1}{c}{416} & \cellcolor[rgb]{ .937,  .937,  .937}84.21 & \cellcolor[rgb]{ .937,  .937,  .937}77.38 & \cellcolor[rgb]{ .937,  .937,  .937}73.26 & \cellcolor[rgb]{ .937,  .937,  .937}9.46 \\
          & \multicolumn{1}{c}{Other} & 87.38 & 75.20 & 77.54 & 9.97 \\
    \bottomrule
    \end{tabular}
            }\\[6pt]
            \subfloat[ReAct Agent]{
\begin{tabular}{cp{3.5em}cccc}
    \toprule
    \multicolumn{1}{p{2.1em}}{\textbf{Model}} & \multicolumn{1}{c}{\textbf{CWE-ID}} 
    & \multicolumn{1}{c}{\textbf{PRE (\%)}} & \multicolumn{1}{c}{\textbf{ACC (\%)}} 
    & \multicolumn{1}{c}{\textbf{F1 (\%)}} & \multicolumn{1}{c}{\textbf{FPR (\%)}} \\
    \midrule
    \multicolumn{1}{c}{\multirow{11}[4]{*}{\makecell{ReAct Agent \\ (GPT-4o)}}} 
    & \multicolumn{1}{c}{119} & 82.34 & 83.66 & 79.06 & 11.12 \\
          & \multicolumn{1}{c}{20} & \cellcolor[rgb]{ .937,  .937,  .937}\textbf{83.62} & \cellcolor[rgb]{ .937,  .937,  .937}79.39 & \cellcolor[rgb]{ .937,  .937,  .937}76.48 & \cellcolor[rgb]{ .937,  .937,  .937}10.65 \\
          & \multicolumn{1}{c}{399} & 79.58 & 77.46 & 69.44 & 11.58 \\
          & \multicolumn{1}{c}{200} & \cellcolor[rgb]{ .937,  .937,  .937}69.98 & \cellcolor[rgb]{ .937,  .937,  .937}80.73 & \cellcolor[rgb]{ .937,  .937,  .937}77.22 & \cellcolor[rgb]{ .937,  .937,  .937}10.24 \\
          & \multicolumn{1}{c}{125} & 74.22 & 80.91 & 77.32 & \textbf{9.85} \\
          & \multicolumn{1}{c}{264} & \cellcolor[rgb]{ .937,  .937,  .937}80.52 & \cellcolor[rgb]{ .937,  .937,  .937}\textbf{88.92} & \cellcolor[rgb]{ .937,  .937,  .937}\textbf{86.23} & \cellcolor[rgb]{ .937,  .937,  .937}13.84 \\
          & \multicolumn{1}{c}{189} & 81.18 & 81.22 & 80.49 & 10.67 \\
          & \multicolumn{1}{c}{476} & \cellcolor[rgb]{ .937,  .937,  .937}72.55 & \cellcolor[rgb]{ .937,  .937,  .937}74.43 & \cellcolor[rgb]{ .937,  .937,  .937}73.67 & \cellcolor[rgb]{ .937,  .937,  .937}10.48 \\
          & \multicolumn{1}{c}{190} & 77.47 & 84.83 & 79.41 & 12.45 \\
          & \multicolumn{1}{c}{416} & \cellcolor[rgb]{ .937,  .937,  .937}77.38 & \cellcolor[rgb]{ .937,  .937,  .937}76.72 & \cellcolor[rgb]{ .937,  .937,  .937}74.98 & \cellcolor[rgb]{ .937,  .937,  .937}10.56 \\
          & \multicolumn{1}{c}{Other} & 80.55 & 74.54 & 79.26 & 11.07 \\
    \bottomrule
    \end{tabular}
            }
        \end{tabular}
    }
    \label{tab:CEW result}
\end{table}

Table \ref{tab:CWEtypes} in Section \ref{type} presents the distribution of various CWE types along with their corresponding quantities. Notably, for this RQ, the dataset exclusively comprises security patches, with non-security patches excluded. To assess the effectiveness of our methods in identifying different types of security patches, we select the best-performing model from each method for comparison, as presented in Table \ref{tab:CEW result}.

Overall, our analysis reveals that all evaluated methods are effective across the ten analyzed vulnerability types. Specifically, the Plain LLM achieves average precision, accuracy, F1 score, and FPR of 86.98\%, 82.95\%, 78.12\%, and 9.18\%, respectively. The Data-Aug LLM achieves average precision, accuracy, F1 score, and FPR of 84.96\%, 80.92\%, 75.88\%, and 10.04\%, respectively. Similarly, the React Agent achieves average precision, accuracy, F1 score, and FPR of 78.13\%, 80.26\%, 77.6\%, and 11.14\%, respectively. Surprisingly, both the average and individual scores indicate that the Plain LLM consistently outperforms the Data-Aug LLM and React Agent across various vulnerability types in terms of precision, accuracy, and F1 score, while also exhibiting a lower FPR. 

However, when considering the results of RQ1, the Data-Aug LLM demonstrates superior overall performance in the task of security patch detection. This suggests that incorporating additional contextual information can enhance model performance for comprehensive security patch detection. Conversely, for single-category security patch detection, commercial LLMs alone are capable of achieving excellent performance, where the inclusion of additional contextual information or iterative reasoning may, in contrast, diminish their effectiveness.

\begin{figure}[!htbp]
    \centering
    \includegraphics[width=3.5in]{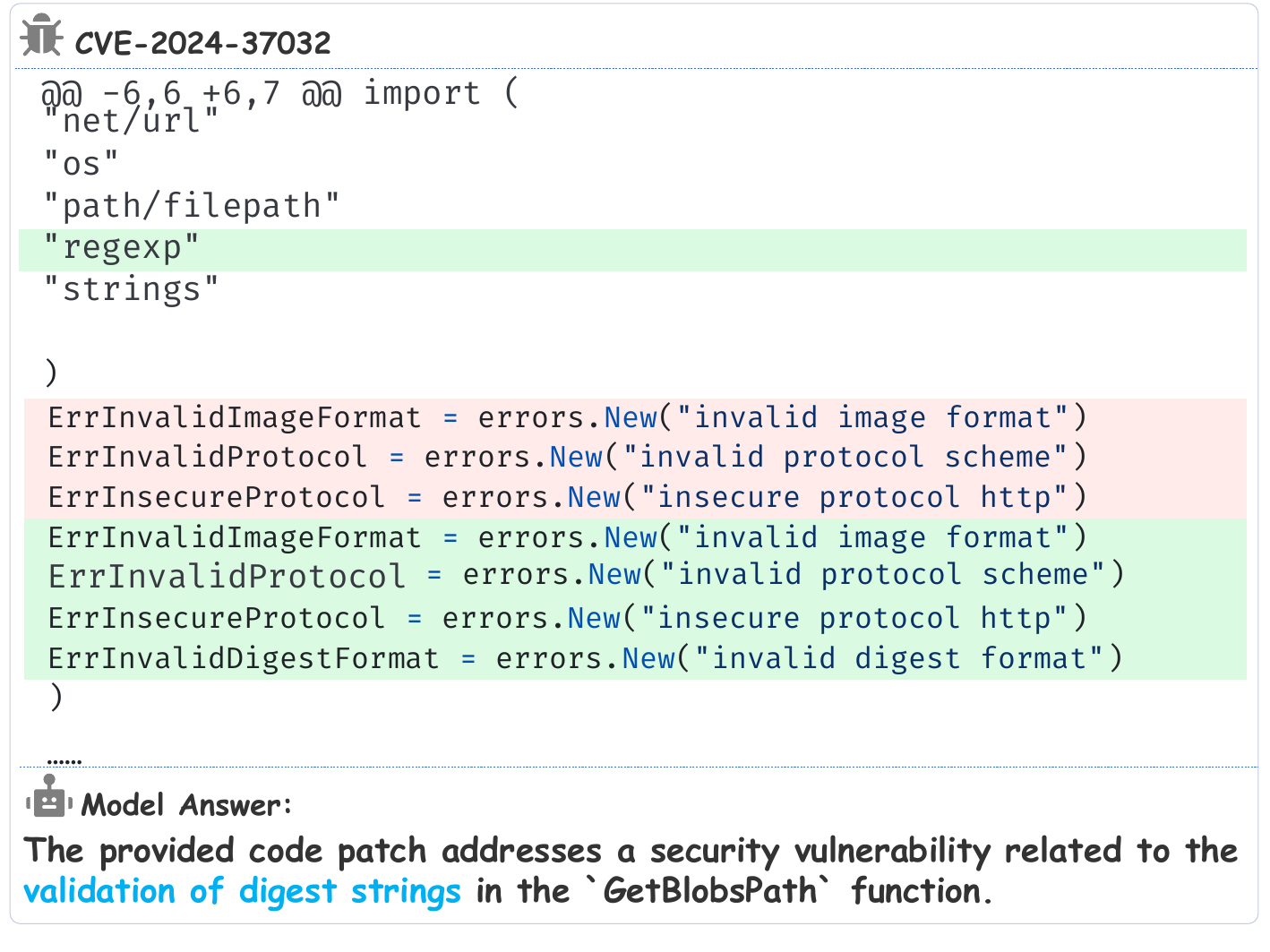}
    \caption{An example of security patch detection results for CVE-2024-37032.}
    \label{fig:example_result}
\end{figure}

Further analysis reveals that all evaluated methods achieve their highest precision in the CWE-20 (Input Validation) category. Specifically, the Plain LLM achieves 92.32\%, the Data-Aug LLM achieves 90.45\%, and the React Agent achieves 83.62\%. Figure \ref{fig:example_result} illustrates a specific example involving CVE-2024-37032, which belongs to the CWE-20 category. The GPT-4o model under the Plain LLM successfully identified the absence of digest string validation and provided a reasonable explanation, confirming that this is a security patch that addresses a security vulnerability in the string validation process.

\begin{figure}[!htbp]
    \centering
    \includegraphics[width=3.5in]{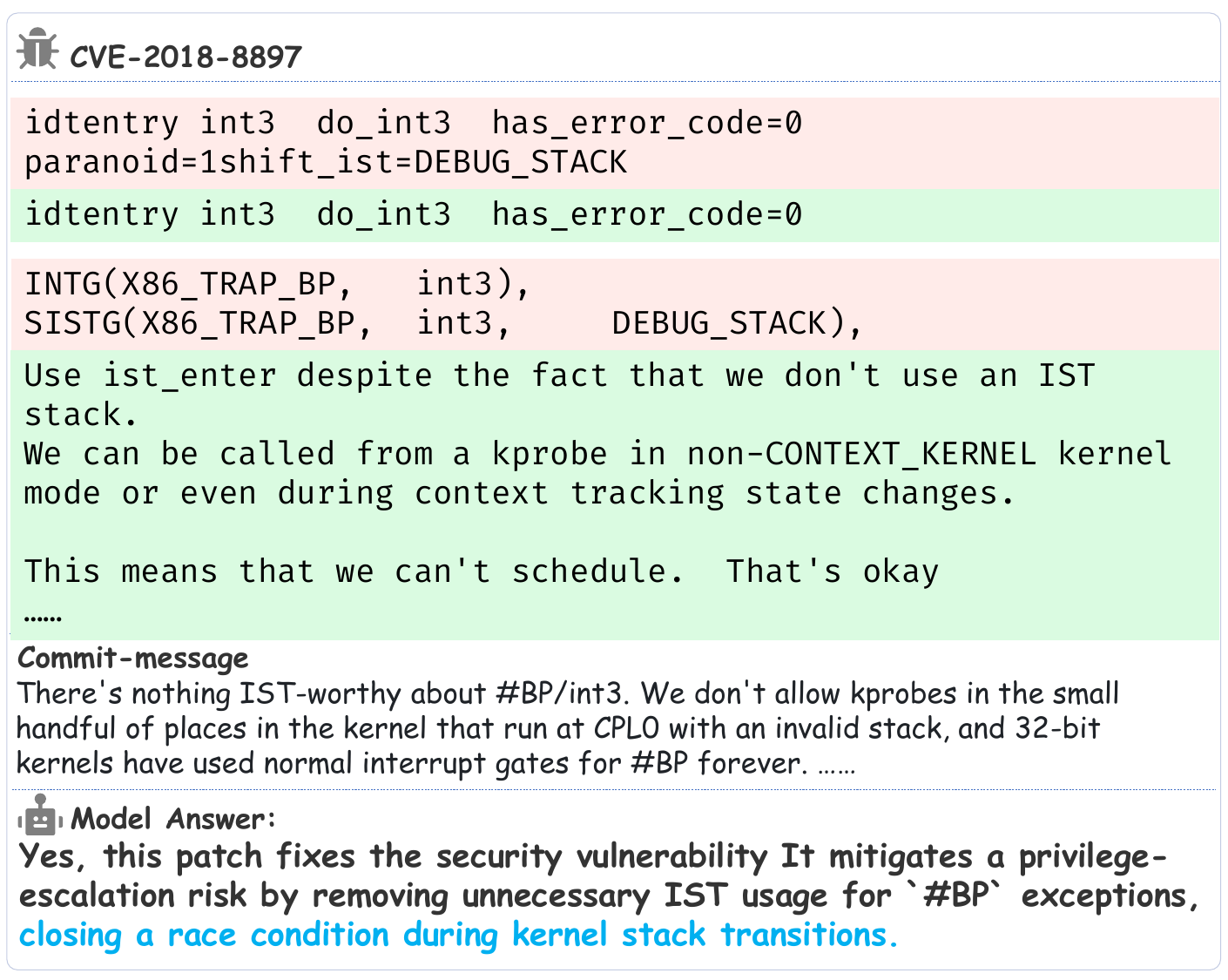}
    \caption{An example of security patch detection results for CVE-2018-8897.}
    \label{fig:example_result2}
\end{figure}

Additionally, our analysis shows that all evaluated methods achieve their highest accuracy and F1 score in the CWE-264 (Authorization Management) category. Specifically, the Plain LLM achieves an accuracy of 91.37\% and an F1 score of 86.88\%, the Data-Aug LLM achieves 89.58\% and 84.51\%, and the React Agent achieves 88.92\% and 86.23\%, respectively. Figure \ref{fig:example_result2} provides an example of CVE-2018-8897, where the Deepseek-R1 model under the Data-Aug method accurately identified the security patch. By analyzing the commit message, the model inferred that the patch removed redundant dependencies on the Interrupt Stack Table (IST), thereby effectively addressing potential race conditions during kernel stack switching and mitigating the risk of privilege escalation. These findings highlight the model's capability for effective reasoning based on contextual information.

Finally, we observe that all evaluated methods achieve their lowest FPR in the CWE-476 (NULL Pointer Dereference) and CWE-125 (Out-of-Bounds Read) categories. Specifically, the Plain LLM achieves an FPR of 8.02\%, the Data-Aug LLM achieves 8.75\%, and the React Agent achieves 9.85\%.

\begin{tcolorbox}[answerbox]
  \textbf{Finding 4:} The evaluation demonstrates that all methods effectively identify security patches across various CWE types, with the Plain LLM consistently outperforming the Data-Aug LLM and React Agent in terms of precision, accuracy, F1 score, and FPR. Notably, all evaluated methods achieve their highest precision in the CWE-20 (Input Validation) category, their highest accuracy and F1 score in the CWE-264 (Authorization Management) category, and their lowest FPR in the CWE-476 (NULL Pointer Dereference) and CWE-125 (Out-of-Bounds Read) categories.
\end{tcolorbox}

\subsection{RQ5: How do different prompt engineering strategies influence the effectiveness of LLMs and LLM-based agents?}

\begin{figure}[!htbp]
    \centering
    \includegraphics[width=3.0in]{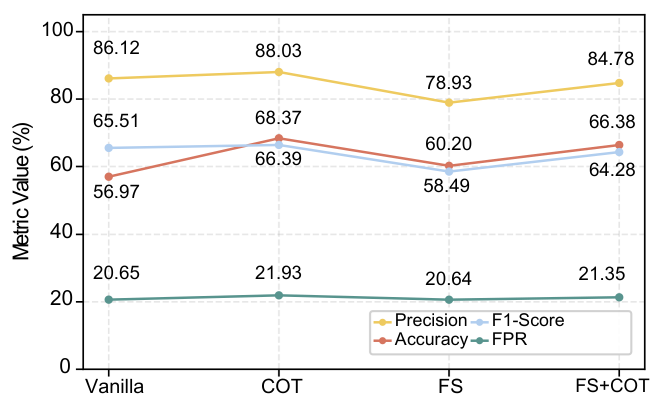}
    \caption{The precision, accuracy, F1 score, and FPR of LLMs (GPT-4o) under the Plain LLM method with various prompting strategies.}
    \label{fig:plain prompt result}
\end{figure}

\begin{figure}[!tbp]
    \centering
    \includegraphics[width=3.0in]{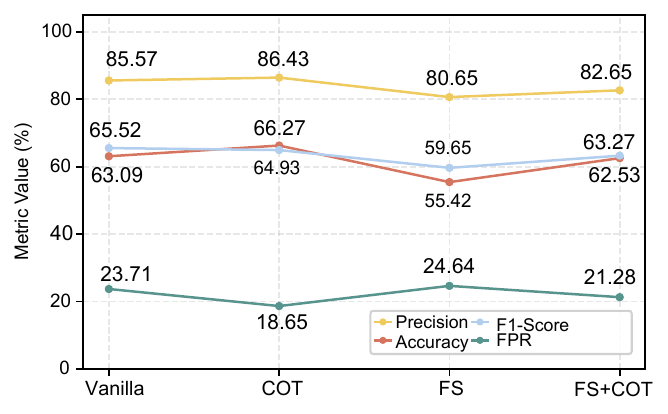}
    \caption{The precision, accuracy, F1 score, and FPR of LLMs (DeepSeek-R1) under the Data-Aug LLM method with various prompting strategies.}
    \label{fig:data-aug prompt result}
\end{figure}

\begin{figure}[!htbp]
    \centering
    \includegraphics[width=3.0in]{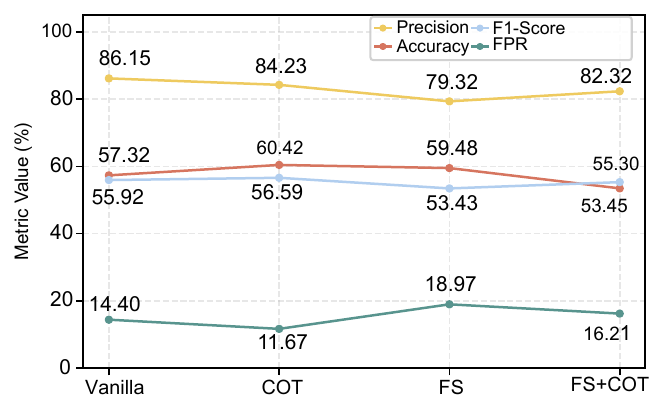}
    \caption{The precision, accuracy, F1 score, and FPR of LLMs (GPT-4o) under the React Agent method with various prompting strategies.}
    \label{fig:agent prompt result}
\end{figure}

We investigated the effects of three prompting strategies on security patch detection across multiple experimental paradigms. For each paradigm, we selected the highest-performing model. Figures \ref{fig:plain prompt result}, \ref{fig:data-aug prompt result}, and \ref{fig:agent prompt result} present the detailed results.

Among the evaluated methods, the CoT prompting strategy demonstrates the most significant improvement in performance. This could be due to the step-by-step reasoning process enabled by CoT, which allows the model to focus on critical semantic and structural features at each step, thereby effectively identifying key security patterns in patches and reducing false positive rates. Specifically, in Plain LLM, the CoT prompting strategy achieves a precision of 88.03\%, outperforming the performance of Vanilla, FS, and FS+CoT prompting strategies by ranging from 1.91\% to 9.1\%. For Data-Aug LLM, the CoT prompting strategy yields a precision of 86.43\%, which is 0.86\% to 5.78\% higher than that of Vanilla, FS, and FS+CoT strategies. This improvement may be partially result from the benefits of data augmentation, which also enhances the performance of other prompting strategies. However, in the ReAct Agent, the precision achieved with the CoT prompting strategy is 84.23\%, compared to 86.15\% with Vanilla. Given that ReAct Agent inherently employs an iterative reasoning process, this suggests that the integration of the CoT strategy may, in certain cases, be less effective than Vanilla strategy.

In contrast, the benefits of the FS strategy are quite limited, with its performance generally lagging behind across most metrics. For Plain LLM, Data-Aug LLM, and ReAct Agent, the precision of models employing the FS strategy is 7.19\%, 4.92\%, and 6.83\% lower than that of the Vanilla strategy, respectively. The decline in performance may be attributed to the inclusion of a small number of samples, which can inadvertently introduce noise or task-irrelevant information, thereby impairing the model's performance. The hybrid strategy that combines CoT and FS produces results that generally fall between those of the CoT and FS strategies when applied individually. While the hybrid strategy outperforms FS alone, its overall performance remains inferior to that of CoT.

\begin{tcolorbox}[answerbox]
  \textbf{Finding 5:} In most cases, the CoT prompting strategy is the most effective for improving model performance, followed by the hybrid strategy combining CoT and FS. The FS strategy performs the worst, as it tends to introduce noise, resulting in a significant decline in performance.
\end{tcolorbox}

\subsection{RQ6: Does the context window (CW) size affect the overall effectiveness of LLMs and LLM-based agents?}

\begin{figure}[!h]
    \centering
    \subfloat[Plain LLM]{
        \includegraphics[width=3.0in]{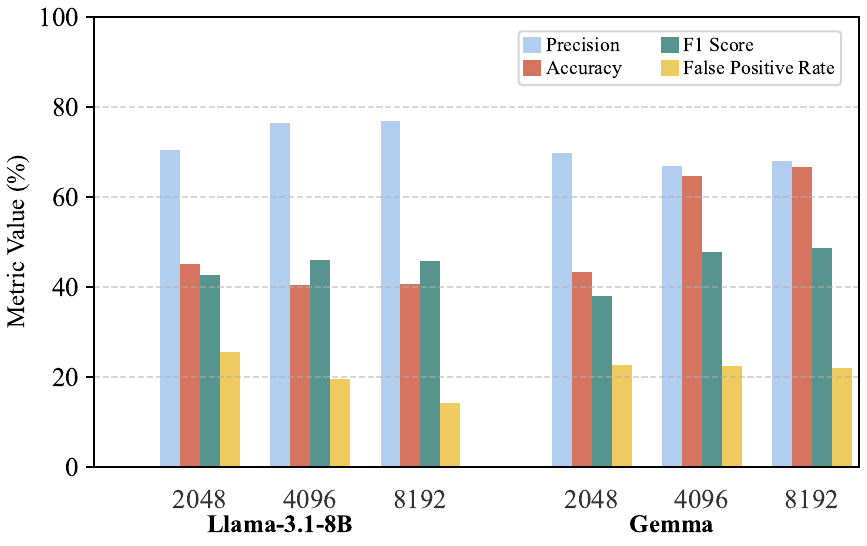}
    }
    \\
    \subfloat[Data-Aug LLM]{
        \includegraphics[width=3.0in]{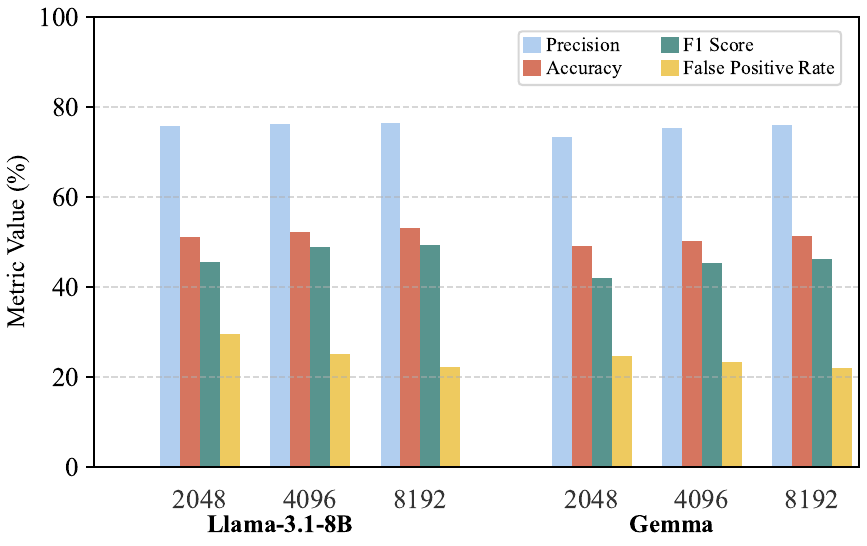}
    }
    \\
    \subfloat[ReAct Agent]{
        \includegraphics[width=3.0in]{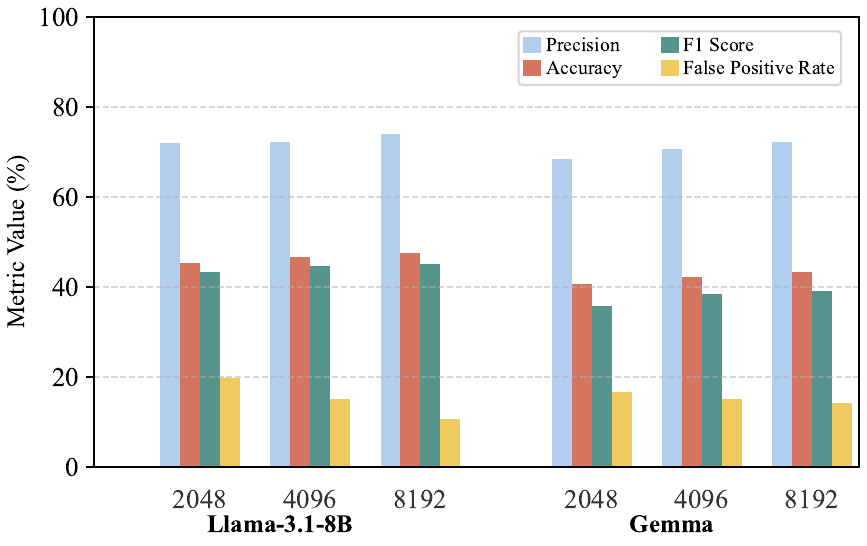}
    }
    \caption{Impact of CW sizes on the performance of open-source LLMs under each evaluated method.}
    \label{fig:CW result}
\end{figure}

Figure \ref{fig:CW result} illustrates that, in most cases, models with larger context window sizes tend to exhibit superior performance across various metrics. For instance, when using LLaMA-3.1, except for the Plain LLM method, the version with a CW of 8,192 tokens achieves higher precision, accuracy, and F1 score, along with a lower false positive rate, compared to the version with a 2,048-token context window. Specifically, under the Data-Aug LLM method, its precision improves from 75.78\% to 76.50\%, accuracy increases from 51.03\% to 53.00\%, and the F1 score rises from 45.64\% to 49.40\%. Meanwhile, the false positive rate decreases by 7.19\%, dropping from 29.47\% to 22.28\%.

\begin{figure}[t]
    \centering
    \includegraphics[width=3.5in]{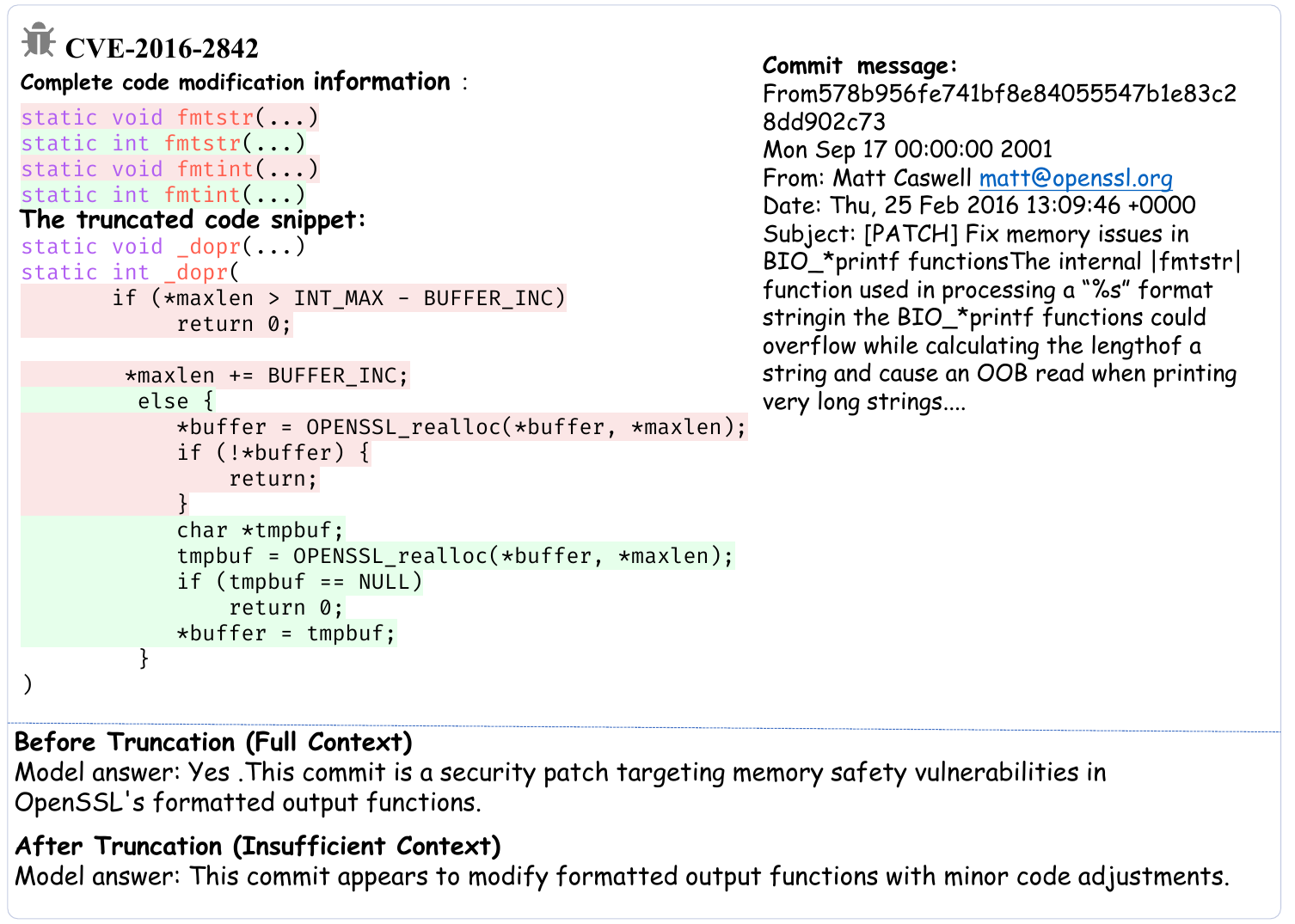}
    \caption{An example of security patch detection results for CVE-2016-2842.}
    \label{fig:CW-ex}
\end{figure}

A notable example is the detection of the CVE-2016-2842 security patch using the Data-Aug LLM method with the LLaMA-3.1 model, as shown in Figure \ref{fig:CW-ex}. In this case, the total length of code changes and commit messages exceeded 16,000 characters (equivalent to over 4,000 tokens). When the model's context window was limited to 2,048 tokens, critical code snippets were truncated, leading to the misclassification of the commit as a regular performance update rather than a security patch. However, by extending the context window to 4,096 tokens, the model successfully processed the complete modification and accurately identified the commit as a security patch. This result highlights the critical dependency of LLMs on context window sizes when processing for reasoning about complex software updates.

Moreover, although Gemma-3's overall performance remains lower than that of LLaMA-3.1, it exhibits a similar trend. Under the React Agent method, the version of Gemma-3 with an 8,192-token context window achieves a 5.42\% improvement in precision, while accuracy increases from 40.65\% to 43.30\%, F1 score rises from 35.68\% to 39.18\%, and FPR decreases by 2.42 percentage points.

\begin{tcolorbox}[answerbox]
  \textbf{Finding 6:} Models with larger context window sizes generally demonstrate enhanced performance across various metrics, as evidenced by LLaMA-3.1 and Gemma-3.
\end{tcolorbox}

\section{Discussion}\label{discuss}
In this section, we discuss the implications of this study and provide practical recommendations for researchers and practitioners.

\textbf{Findings in RQ1} reveal that leveraging data augmentation enables the Data-Aug LLM to achieve superior overall performance in the security patch detection task, while the ReAct Agent demonstrates the lowest false positive rate (FPR) due to its iterative reasoning and action mechanism. In this regard, researchers could explore hybrid models that balance the trade-off between utilizing additional contextual information and mitigating the noise it introduces. For instance, combining data augmentation techniques and an iterative reasoning process for enhanced performance. Moreover, researchers should develop and implement robust validation steps or post-processing mechanisms to address the higher false positive rates caused by the noise in augmented data, thereby ensuring the enhancement of data augmentation in LLMs.

For practitioners, the selection of methods should be guided by specific scenarios. In scenarios that are highly sensitive to false positives, the ReAct Agent is recommended due to its balance between detection precision and false positive rates. Conversely, in scenarios of broader vulnerability coverage, the Data-Aug LLM method is more suitable. Users adopting automated detection tools should carefully evaluate the trade-off between detection accuracy and false positives to ensure alignment with their specific needs.

\textbf{Findings in RQ2} indicate that commercial LLMs, such as GPT-4o, GPT-4o-mini, and DeepSeek-R1, consistently outperform open-source LLMs like Llama-3.1 and Gemma-3 across the Plain LLM, Data-Aug LLM, and ReAct Agent methods. For patch management practitioners, it is recommended to prioritize closed-source models in scenarios with abundant resources, as they can provide more reliable detection performance. Conversely, in scenarios with limited resources, fine-tuning open-source models with task-specific datasets may be a practical strategy to enhance their effectiveness in real-world applications, enabling them to serve as viable alternatives to commercial models.

\textbf{Findings in RQ3} unveil that, compared to traditional feature-based or graph-structured approaches, the LLM-based ReAct Agent method demonstrates superior detection performance while maintaining a relatively low false positive rate. Future research could explore the deep integration of LLMs with structured code representations, such as graph learning or abstract syntax tree (AST), to further enhance detection capabilities. For practitioners in patch management, incorporating LLM-based methods into existing security patch detection workflows could leverage the strengths of both approaches. For instance, a dual-detection mechanism that combines traditional approaches with LLM-based approaches may facilitate earlier and more accurate identification of security patches.

\textbf{Findings from RQ4} uncover that, although all methods are generally effective in identifying security patches across various CWE categories, their detection performance varies among different CWE types. Specifically, the evaluated methods achieve the highest precision in the CWE-20 (Input Validation) category, the highest accuracy and F1 score in the CWE-264 (Authorization Management) category, and the lowest false positive rate in the CWE-476 (Null Pointer Dereference) and CWE-125 (Out-of-Bounds Read) categories. These differences reflect the inherent complexity of certain vulnerability types, which impacts detection performance, as well as the issue of class imbalance present in existing datasets.

For researchers, future efforts should focus on developing CWE-aware detection methods capable of capturing the heterogeneity of vulnerability categories more effectively. For example, leveraging a Mixture of Experts (MoE) architecture, refining prompt strategies for complex CWE types, and employing data augmentation or transfer learning techniques for rare vulnerabilities could improve overall performance while enabling more precise detection across diverse vulnerability types.

For practitioners, it is essential to recognize that automated detection systems are not infallible, especially in high-risk or safety-critical domains. Such systems may demonstrate superior detection performance for specific vulnerability types, but should not be relied upon as definitive decision-makers. Instead, their outputs should be treated as decision-support tools, complemented by manual review or cross-validation using multiple detection tools to establish a more comprehensive and resilient security assurance workflow.

\textbf{Findings from RQ5} give a hint that the CoT (Chain-of-Thought) prompting strategy is, in most cases, the most effective approach for improving model performance. This is followed by the hybrid strategy that combines CoT and FS (Few-Shot) approaches. The FS strategy performs the worst, as it is prone to introducing noise, which significantly degrades its effectiveness. Researchers are encouraged to leverage the CoT strategy for security patch detection tasks while further investigating ways to optimize prompt templates and refine example selection strategies to improve the effectiveness of FS methods.

For practitioners in patch management, the CoT prompting strategy should be prioritized in practical applications due to its superior performance. If the FS method is to be utilized, it is imperative to adopt appropriate example selection techniques and carefully validate the selected examples to ensure their accuracy and representativeness.

\textbf{Findings in RQ6} demonstrate that the size of the context window significantly impacts the performance of open-source LLMs. Expanding the context window sizes consistently enhances model performance, particularly in scenarios involving complex code modifications. However, when the context window size is constrained, researchers should focus on developing techniques to effectively extract critical information from lengthy code sequences, such as leveraging hierarchical attention mechanisms, to enhance the model's ability to utilize extended contexts.

For practitioners in patch management, it is advisable to allocate the largest possible context window to open-source LLMs within the limits of available computational resources. This approach represents a straightforward yet effective engineering practice that can substantially improve detection performance.

\section{THREATS TO VALIDITY}\label{threats}
\noindent\textbf{Threats to Internal Validity.} The first threat to internal validity concerns the selection of LLMs. To address this threat, we included both closed-source and open-source models into our comparison, selecting widely recognized and commonly used models from each category to ensure the reliability of the comparative results. The second threat pertains to the comparison with baselines, specifically the reliability of reproducing baseline results. To mitigate this, we faithfully utilized the official implementations provided for each baseline during reproduction. All selected baseline methods were sourced from publicly available GitHub repositories. However, as the output granularity of certain baselines was not aligned with the binary outputs of our security patch detection task, we adapted their outputs to match the binary format, enabling a more meaningful comparison. Notably, these adaptations may introduce slight deviations from the original results reported in the literature.

\noindent\textbf{Threats to External Validity.} The threat to external validity relates to the generalizability of the research findings. A key limitation lies in the scope of the PatchDB dataset, which focuses exclusively on security patches written in C/C++. Consequently, our conclusions may not be fully generalizable to other programming languages or the broader software ecosystem. To alleviate this limitation, future work aim to extend the proposed approach to incorporate a more diverse range of programming languages and vulnerability datasets, thereby enhancing its generalizability and applicability.

\noindent\textbf{Threats to Construct Validity.} The threat to construct validity lies in whether the evaluation metrics used to measure performance are sufficiently comprehensive. To address this, we adopted four widely accepted metrics in vulnerability detection research—accuracy, precision, recall, and F1 score. We believe that these metrics provide a holistic assessment of the model's effectiveness and contribute to enhancing the robustness and reliability of our experimental evaluation.

\section{Conclusion And Future Work}\label{con}
This paper presents the first systematic study evaluating LLMs and LLM-based React Agents for security patch detection. We examined the effectiveness of three methods: Plain LLM, Data-Aug LLM, and React Agent (RQ1), assessed the performance of various LLMs within these methods (RQ2), conducted a comprehensive comparison of the studied methods against existing baselines (RQ3), analyzed the performance of these methods across different types of security vulnerabilities (RQ4), investigated the impact of various prompting strategies (RQ5), and explored the influence of context window sizes on detection performance (RQ6). 

Consequently, our experimental findings demonstrate that Data-Aug LLM achieves superior overall performance compared to the other evaluated methods, while the ReAct Agent achieves the lowest false positive rate due to its reasoning and action process. Additionally, commercial LLMs, such as GPT-4o, GPT-4o-mini, and DeepSeek-R1, consistently outperform open-source LLMs like Llama-3.1 and Gemma-3 across all evaluated methods. All methods achieve their highest precision in the CWE-20 (Input Validation) category, and the CoT prompting strategy proves to be the most effective for improving model performance in most cases. Moreover, open-source LLMs with larger context window sizes generally exhibit enhanced performance across various metrics.

In the future, our research aims to expand to datasets encompassing a wider range of programming languages, with a focus on developing methods that achieve optimal overall performance while maintaining strong detection capabilities for complex and rare CWE types.

\balance
\bibliographystyle{IEEEtran}
\bibliography{ref.bib}

@misc{blackduck,
    author       = "{Black Duck}",
    title        = "Application Security Software (AppSec)",
    year         = "2025",
    howpublished = "\url{https://www.blackduck.com/en-us.html}",
    note         = "Retrieved from the World Wide Web"
}

@inproceedings{wang2019detecting,
  title={Detecting" 0-day" vulnerability: An empirical study of secret security patch in OSS},
  author={Wang, Xinda and Sun, Kun and Batcheller, Archer and Jajodia, Sushil},
  booktitle={2019 49th Annual IEEE/IFIP International Conference on Dependable Systems and Networks (DSN)},
  pages={485--492},
  year={2019},
  organization={IEEE}
}

@inproceedings{wang2021patchrnn,
  title={Patchrnn: A deep learning-based system for security patch identification},
  author={Wang, Xinda and Wang, Shu and Feng, Pengbin and Sun, Kun and Jajodia, Sushil and Benchaaboun, Sanae and Geck, Frank},
  booktitle={MILCOM 2021-2021 IEEE Military Communications Conference (MILCOM)},
  pages={595--600},
  year={2021},
  organization={IEEE}
}

@article{tamjidyamcholo2022subjectivity,
  title={Subjectivity reduction of qualitative approach in information security risk analysis},
  author={TamjidYamcholo, Alireza and Toloie Eshlaghy, Abbas},
  journal={Journal of System Management},
  volume={1},
  number={1},
  pages={145},
  year={2022},
  publisher={Islamic Azad University}
}

@inproceedings{li2017large,
  title={A large-scale empirical study of security patches},
  author={Li, Frank and Paxson, Vern},
  booktitle={Proceedings of the 2017 ACM SIGSAC Conference on Computer and Communications Security},
  pages={2201--2215},
  year={2017}
}

@inproceedings{mirhosseini2017can,
  title={Can automated pull requests encourage software developers to upgrade out-of-date dependencies?},
  author={Mirhosseini, Samim and Parnin, Chris},
  booktitle={2017 32nd IEEE/ACM international conference on automated software engineering (ASE)},
  pages={84--94},
  year={2017},
  organization={IEEE}
}

@inproceedings{thung2012would,
  title={When would this bug get reported?},
  author={Thung, Ferdian and Lo, David and Jiang, Lingxiao and Rahman, Foyzur and Devanbu, Premkumar T and others},
  booktitle={2012 28th IEEE International Conference on Software Maintenance (ICSM)},
  pages={420--429},
  year={2012},
  organization={IEEE}
}

@article{cheng2022bug,
  title={How about bug-triggering paths?-understanding and characterizing learning-based vulnerability detectors},
  author={Cheng, Xiao and Nie, Xu and Li, Ningke and Wang, Haoyu and Zheng, Zheng and Sui, Yulei},
  journal={IEEE Transactions on Dependable and Secure Computing},
  volume={21},
  number={2},
  pages={542--558},
  year={2022},
  publisher={IEEE}
}

@article{zhang2023multi,
  title={Multi-task framework of precipitation nowcasting},
  author={Zhang, Zheng and Luo, Chuyao and Zhang, Baoquan and Jiang, Hao and Zhang, Bowen},
  journal={CAAI Transactions on Intelligence Technology},
  volume={8},
  number={4},
  pages={1350--1363},
  year={2023},
  publisher={Wiley Online Library}
}

@article{liu2024survey,
  title={A survey on federated learning: a perspective from multi-party computation},
  author={Liu, Fengxia and Zheng, Zhiming and Shi, Yexuan and Tong, Yongxin and Zhang, Yi},
  journal={Frontiers of Computer Science},
  volume={18},
  number={1},
  pages={181336},
  year={2024},
  publisher={Springer}
}

@article{brown2020language,
  title={Language models are few-shot learners},
  author={Brown, Tom and Mann, Benjamin and Ryder, Nick and Subbiah, Melanie and Kaplan, Jared D and Dhariwal, Prafulla and Neelakantan, Arvind and Shyam, Pranav and Sastry, Girish and Askell, Amanda and others},
  journal={Advances in neural information processing systems},
  volume={33},
  pages={1877--1901},
  year={2020}
}

@inproceedings{wang2021patchdb,
  title={Patchdb: A large-scale security patch dataset},
  author={Wang, Xinda and Wang, Shu and Feng, Pengbin and Sun, Kun and Jajodia, Sushil},
  booktitle={2021 51st Annual IEEE/IFIP International Conference on Dependable Systems and Networks (DSN)},
  pages={149--160},
  year={2021},
  organization={IEEE}
}

@inproceedings{lin2025large,
  title={From large to mammoth: A comparative evaluation of large language models in vulnerability detection},
  author={Lin, Jie and Mohaisen, David},
  booktitle={Proceedings of the 2025 Network and Distributed System Security Symposium (NDSS)},
  year={2025}
}

@article{yang2025context,
  title={Context-enhanced vulnerability detection based on large language model},
  author={Yang, Yixin and Xu, Bowen and Gao, Xiang and Sun, Hailong},
  journal={arXiv preprint arXiv:2504.16877},
  year={2025}
}

@article{yildiz2025benchmarking,
  title={Benchmarking LLMs and LLM-based Agents in Practical Vulnerability Detection for Code Repositories},
  author={Yildiz, Alperen and Teo, Sin G and Lou, Yiling and Feng, Yebo and Wang, Chong and Divakaran, Dinil M},
  journal={arXiv preprint arXiv:2503.03586},
  year={2025}
}

@article{wang2025vulagent,
  title={VulAgent: Hypothesis-Validation based Multi-Agent Vulnerability Detection},
  author={Wang, Ziliang and Li, Ge and Li, Jia and Zhu, Hao and Jin, Zhi},
  journal={arXiv preprint arXiv:2509.11523},
  year={2025}
}

@inproceedings{yu2025preliminary,
  title={A Preliminary Study of Large Language Models for Multilingual Vulnerability Detection},
  author={Yu, Junji and Shu, Honglin and Fu, Michael and Wang, Dong and Tantithamthavorn, Chakkrit and Kamei, Yasutaka and Chen, Junjie},
  booktitle={Proceedings of the 34th ACM SIGSOFT International Symposium on Software Testing and Analysis},
  pages={161--168},
  year={2025}
}

@inproceedings{fu2022linevul,
  title={Linevul: A transformer-based line-level vulnerability prediction},
  author={Fu, Michael and Tantithamthavorn, Chakkrit},
  booktitle={Proceedings of the 19th International Conference on Mining Software Repositories},
  pages={608--620},
  year={2022}
}

@inproceedings{vaniea2016tales,
  title={Tales of software updates: The process of updating software},
  author={Vaniea, Kami and Rashidi, Yasmeen},
  booktitle={Proceedings of the 2016 chi conference on human factors in computing systems},
  pages={3215--3226},
  year={2016}
}

@inproceedings{huang2019using,
  title={Using safety properties to generate vulnerability patches},
  author={Huang, Zhen and Lie, David and Tan, Gang and Jaeger, Trent},
  booktitle={2019 IEEE symposium on security and privacy (SP)},
  pages={539--554},
  year={2019},
  organization={IEEE}
}

@inproceedings{wu2020precisely,
  title={Precisely characterizing security impact in a flood of patches via symbolic rule comparison},
  author={Wu, Qiushi and He, Yang and McCamant, Stephen and Lu, Kangjie},
  booktitle={The 2020 Annual Network and Distributed System Security Symposium (NDSS'20)},
  year={2020}
}

@inproceedings{tian2012identifying,
  title={Identifying linux bug fixing patches},
  author={Tian, Yuan and Lawall, Julia and Lo, David},
  booktitle={2012 34th international conference on software engineering (ICSE)},
  pages={386--396},
  year={2012},
  organization={IEEE}
}

@inproceedings{xu2020automatic,
  title={Automatic hot patch generation for android kernels},
  author={Xu, Zhengzi and Zhang, Yulong and Zheng, Longri and Xia, Liangzhao and Bao, Chenfu and Wang, Zhi and Liu, Yang},
  booktitle={29th USENIX Security Symposium (USENIX Security 20)},
  pages={2397--2414},
  year={2020}
}

@inproceedings{wang2020machine,
  title={A machine learning approach to classify security patches into vulnerability types},
  author={Wang, Xinda and Wang, Shu and Sun, Kun and Batcheller, Archer and Jajodia, Sushil},
  booktitle={2020 IEEE Conference on Communications and Network Security (CNS)},
  pages={1--9},
  year={2020},
  organization={IEEE}
}

@inproceedings{soto2016deeper,
  title={A deeper look into bug fixes: patterns, replacements, deletions, and additions},
  author={Soto, Mauricio and Thung, Ferdian and Wong, Chu-Pan and Le Goues, Claire and Lo, David},
  booktitle={Proceedings of the 13th International Conference on Mining Software Repositories},
  pages={512--515},
  year={2016}
}

@inproceedings{corley2011recovering,
  title={Recovering traceability links between source code and fixed bugs via patch analysis},
  author={Corley, Christopher S and Kraft, Nicholas A and Etzkorn, Letha H and Lukins, Stacy K},
  booktitle={Proceedings of the 6th international workshop on traceability in emerging forms of software engineering},
  pages={31--37},
  year={2011}
}

@article{zhou2021spi,
  title={Spi: Automated identification of security patches via commits},
  author={Zhou, Yaqin and Siow, Jing Kai and Wang, Chenyu and Liu, Shangqing and Liu, Yang},
  journal={ACM Transactions on Software Engineering and Methodology (TOSEM)},
  volume={31},
  number={1},
  pages={1--27},
  year={2021},
  publisher={ACM New York, NY}
}

@inproceedings{nguyen2022vulcurator,
  title={Vulcurator: a vulnerability-fixing commit detector},
  author={Nguyen, Truong Giang and Le-Cong, Thanh and Kang, Hong Jin and Le, Xuan-Bach D and Lo, David},
  booktitle={Proceedings of the 30th ACM Joint European Software Engineering Conference and Symposium on the Foundations of Software Engineering},
  pages={1726--1730},
  year={2022}
}

@article{wu2022enhancing,
  title={Enhancing security patch identification by capturing structures in commits},
  author={Wu, Bozhi and Liu, Shangqing and Feng, Ruitao and Xie, Xiaofei and Siow, Jingkai and Lin, Shang-Wei},
  journal={IEEE Transactions on Dependable and Secure Computing},
  year={2022},
  publisher={IEEE}
}

@inproceedings{zhou2023colefunda,
  title={Colefunda: Explainable silent vulnerability fix identification},
  author={Zhou, Jiayuan and Pacheco, Michael and Chen, Jinfu and Hu, Xing and Xia, Xin and Lo, David and Hassan, Ahmed E},
  booktitle={2023 IEEE/ACM 45th International Conference on Software Engineering (ICSE)},
  pages={2565--2577},
  year={2023},
  organization={IEEE}
}

@inproceedings{zuo2023commit,
  title={Commit message can help: security patch detection in open source software via transformer},
  author={Zuo, Fei and Zhang, Xin and Song, Yuqi and Rhee, Junghwan and Fu, Jicheng},
  booktitle={2023 IEEE/ACIS 21st International Conference on Software Engineering Research, Management and Applications (SERA)},
  pages={345--351},
  year={2023},
  organization={IEEE}
}

@inproceedings{wang2023graphspd,
  title={Graphspd: Graph-based security patch detection with enriched code semantics},
  author={Wang, Shu and Wang, Xinda and Sun, Kun and Jajodia, Sushil and Wang, Haining and Li, Qi},
  booktitle={2023 IEEE Symposium on Security and Privacy (SP)},
  pages={2409--2426},
  year={2023},
  organization={IEEE}
}

@inproceedings{zhou2021finding,
  title={Finding a needle in a haystack: Automated mining of silent vulnerability fixes},
  author={Zhou, Jiayuan and Pacheco, Michael and Wan, Zhiyuan and Xia, Xin and Lo, David and Wang, Yuan and Hassan, Ahmed E},
  booktitle={2021 36th IEEE/ACM International Conference on Automated Software Engineering (ASE)},
  pages={705--716},
  year={2021},
  organization={IEEE}
}

@inproceedings{li2024patchfinder,
  title={Patchfinder: A two-phase approach to security patch tracing for disclosed vulnerabilities in open-source software},
  author={Li, Kaixuan and Zhang, Jian and Chen, Sen and Liu, Han and Liu, Yang and Chen, Yixiang},
  booktitle={Proceedings of the 33rd ACM SIGSOFT International Symposium on Software Testing and Analysis},
  pages={590--602},
  year={2024}
}

@article{li2025they,
  title={What Do They Fix? LLM-Aided Categorization of Security Patches for Critical Memory Bugs},
  author={Li, Xingyu and Pu, Juefei and Wu, Yifan and Zou, Xiaochen and Zhu, Shitong and Wu, Qiushi and Zhang, Zheng and Hsu, Joshua and Dong, Yue and Qian, Zhiyun and others},
  journal={arXiv preprint arXiv:2509.22796},
  year={2025}
}

@article{li2025empirical,
  title={Empirical study of code large language models for binary security patch detection},
  author={Li, Qingyuan and Li, Binchang and Gao, Cuiyun and Gao, Shuzheng and Li, Zongjie},
  journal={arXiv preprint arXiv:2509.06052},
  year={2025}
}

@inproceedings{wang2023rap,
  title={Rap-gen: Retrieval-augmented patch generation with codet5 for automatic program repair},
  author={Wang, Weishi and Wang, Yue and Joty, Shafiq and Hoi, Steven CH},
  booktitle={Proceedings of the 31st ACM Joint European Software Engineering Conference and Symposium on the Foundations of Software Engineering},
  pages={146--158},
  year={2023}
}

@inproceedings{wang2022vcmatch,
  title={Vcmatch: a ranking-based approach for automatic security patches localization for OSS vulnerabilities},
  author={Wang, Shichao and Zhang, Yun and Bao, Liagfeng and Xia, Xin and Wu, Minghui},
  booktitle={2022 IEEE International Conference on Software Analysis, Evolution and Reengineering (SANER)},
  pages={589--600},
  year={2022},
  organization={IEEE}
}

@article{wen2024repository,
  title={Repository-Level Graph Representation Learning for Enhanced Security Patch Detection},
  author={Wen, Xin-Cheng and Lin, Zirui and Gao, Cuiyun and Zhang, Hongyu and Wang, Yong and Liao, Qing},
  journal={arXiv preprint arXiv:2412.08068},
  year={2024}
}

@article{roziere2023llamacode,
  title={Code llama: Open foundation models for code},
  author={Roziere, Baptiste and Gehring, Jonas and Gloeckle, Fabian and Sootla, Sten and Gat, Itai and Tan, Xiaoqing Ellen and Adi, Yossi and Liu, Jingyu and Sauvestre, Romain and Remez, Tal and others},
  journal={arXiv preprint arXiv:2308.12950},
  year={2023}
}

@inproceedings{luo2024strengthening,
  title={Strengthening supply chain security with fine-grained safe patch identification},
  author={Luo, Changhua and Meng, Wei and Wang, Shuai},
  booktitle={Proceedings of the IEEE/ACM 46th International Conference on Software Engineering},
  pages={1--12},
  year={2024}
}

@inproceedings{lin2024one,
  title={One size does not fit all: Multi-granularity patch generation for better automated program repair},
  author={Lin, Bo and Wang, Shangwen and Wen, Ming and Chen, Liqian and Mao, Xiaoguang},
  booktitle={Proceedings of the 33rd ACM SIGSOFT International Symposium on Software Testing and Analysis},
  pages={1554--1566},
  year={2024}
}

@inproceedings{xie2024unveiling,
  title={Unveiling the Characteristics and Impact of Security Patch Evolution},
  author={Xie, Zifan and Wen, Ming and Wei, Zichao and Jin, Hai},
  booktitle={Proceedings of the 39th IEEE/ACM International Conference on Automated Software Engineering},
  pages={1094--1106},
  year={2024}
}

@article{tian2023best,
  title={The best of both worlds: Combining learned embeddings with engineered features for accurate prediction of correct patches},
  author={Tian, Haoye and Liu, Kui and Li, Yinghua and Kabor{\'e}, Abdoul Kader and Koyuncu, Anil and Habib, Andrew and Li, Li and Wen, Junhao and Klein, Jacques and Bissyand{\'e}, Tegawend{\'e} F},
  journal={ACM Transactions on Software Engineering and Methodology},
  volume={32},
  number={4},
  pages={1--34},
  year={2023},
  publisher={ACM New York, NY, USA}
}

@article{tang2023just,
  title={Just-in-Time Detection of Silent Security Patches},
  author={Tang, Xunzhu and Kim, Kisub and Ezzini, Saad and Song, Yewei and Tian, Haoye and Klein, Jacques and Bissyande, Tegawende},
  journal={ACM Transactions on Software Engineering and Methodology},
  year={2023},
  publisher={ACM New York, NY}
}

@techreport{Synopsys2024OSSRA,
  author      = {{Synopsys}},
  title       = {2024 Open Source Security and Risk Analysis Report},
  institution = {Synopsys, Inc.},
  year        = {2024},
  type        = {Report},
  note        = {[Online]. Available: \url{https://www.synopsys.com/content/dam/synopsys/sig-assets/reports/rep-ossra-2024.pdf}},
  url         = {https://www.synopsys.com/content/dam/synopsys/sig-assets/reports/rep-ossra-2024.pdf},
  urldate     = {2025-09-23}
}

@article{elman1990finding,
  title={Finding structure in time},
  author={Elman, Jeffrey L},
  journal={Cognitive science},
  volume={14},
  number={2},
  pages={179--211},
  year={1990},
  publisher={Wiley Online Library}
}

@inproceedings{huo2020control,
  title={Control flow graph embedding based on multi-instance decomposition for bug localization},
  author={Huo, Xuan and Li, Ming and Zhou, Zhi-Hua},
  booktitle={Proceedings of the AAAI conference on artificial intelligence},
  volume={34},
  number={04},
  pages={4223--4230},
  year={2020}
}

@inproceedings{cummins2021programl,
  title={Programl: A graph-based program representation for data flow analysis and compiler optimizations},
  author={Cummins, Chris and Fisches, Zacharias V and Ben-Nun, Tal and Hoefler, Torsten and O’Boyle, Michael FP and Leather, Hugh},
  booktitle={International Conference on Machine Learning},
  pages={2244--2253},
  year={2021},
  organization={PMLR}
}

@article{li2015gated,
  title={Gated graph sequence neural networks},
  author={Li, Yujia and Tarlow, Daniel and Brockschmidt, Marc and Zemel, Richard},
  journal={arXiv preprint arXiv:1511.05493},
  year={2015}
}

@article{huang2024can,
  title={Can large language models identify authorship?},
  author={Huang, Baixiang and Chen, Canyu and Shu, Kai},
  journal={arXiv preprint arXiv:2403.08213},
  year={2024}
}

@article{zhang2025llm,
  title={Llm hallucinations in practical code generation: Phenomena, mechanism, and mitigation},
  author={Zhang, Ziyao and Wang, Chong and Wang, Yanlin and Shi, Ensheng and Ma, Yuchi and Zhong, Wanjun and Chen, Jiachi and Mao, Mingzhi and Zheng, Zibin},
  journal={Proceedings of the ACM on Software Engineering},
  volume={2},
  number={ISSTA},
  pages={481--503},
  year={2025},
  publisher={ACM New York, NY, USA}
}

@article{pan2025codecor,
  title={CodeCoR: An LLM-based self-reflective multi-agent framework for code generation},
  author={Pan, Ruwei and Zhang, Hongyu and Liu, Chao},
  journal={arXiv preprint arXiv:2501.07811},
  year={2025}
}

@article{wang2025teaching,
  title={Teaching code llms to use autocompletion tools in repository-level code generation},
  author={Wang, Chong and Zhang, Jian and Feng, Yebo and Li, Tianlin and Sun, Weisong and Liu, Yang and Peng, Xin},
  journal={ACM Transactions on Software Engineering and Methodology},
  volume={34},
  number={7},
  pages={1--27},
  year={2025},
  publisher={ACM New York, NY}
}

@article{li2025context,
  title={Context-aware prompting for LLM-based program repair},
  author={Li, Yingling and Cai, Muxin and Chen, Junjie and Xu, Yang and Huang, Lei and Li, Jianping},
  journal={Automated Software Engineering},
  volume={32},
  number={2},
  pages={42},
  year={2025},
  publisher={Springer}
}

@article{huang2025comprehensive,
  title={Comprehensive Fine-Tuning Large Language Models of Code for Automated Program Repair},
  author={Huang, Kai and Zhang, Jian and Bao, Xinlei and Wang, Xu and Liu, Yang},
  journal={IEEE Transactions on Software Engineering},
  year={2025},
  publisher={IEEE}
}

@article{kong2025demystifying,
  title={Demystifying Memorization in LLM-Based Program Repair via a General Hypothesis Testing Framework},
  author={Kong, Jiaolong and Xie, Xiaofei and Liu, Shangqing},
  journal={Proceedings of the ACM on Software Engineering},
  volume={2},
  number={FSE},
  pages={2712--2734},
  year={2025},
  publisher={ACM New York, NY, USA}
}

@article{hussain2025vulbinllm,
  title={VulBinLLM: LLM-powered Vulnerability Detection for Stripped Binaries},
  author={Hussain, Nasir and Chen, Haohan and Tran, Chanh and Huang, Philip and Li, Zhuohao and Chugh, Pravir and Chen, William and Kundu, Ashish and Tian, Yuan},
  journal={arXiv preprint arXiv:2505.22010},
  year={2025}
}

@article{ding2025smartguard,
  title={SmartGuard: An LLM-enhanced framework for smart contract vulnerability detection},
  author={Ding, Hao and Liu, Yizhou and Piao, Xuefeng and Song, Huihui and Ji, Zhenzhou},
  journal={Expert Systems with Applications},
  volume={269},
  pages={126479},
  year={2025},
  publisher={Elsevier}
}

@article{tihanyi2025vulnerability,
  title={Vulnerability detection: from formal verification to large language models and hybrid approaches: a comprehensive overview},
  author={Tihanyi, Norbert and Bisztray, Tamas and Ferrag, Mohamed Amine and Cherif, Bilel and Dubniczky, Richard A and Jain, Ridhi and Cordeiro, Lucas C},
  journal={arXiv preprint arXiv:2503.10784},
  year={2025}
}

@article{he2025llm,
  title={LLM-Based Multi-Agent Systems for Software Engineering: Literature Review, Vision, and the Road Ahead},
  author={He, Junda and Treude, Christoph and Lo, David},
  journal={ACM Transactions on Software Engineering and Methodology},
  volume={34},
  number={5},
  pages={1--30},
  year={2025},
  publisher={ACM New York, NY}
}

@article{ataei2025elicitron,
  title={Elicitron: A large language model agent-based simulation framework for design requirements elicitation},
  author={Ataei, Mohammadmehdi and Cheong, Hyunmin and Grandi, Daniele and Wang, Ye and Morris, Nigel and Tessier, Alexander},
  journal={Journal of Computing and Information Science in Engineering},
  volume={25},
  number={2},
  pages={021012},
  year={2025},
  publisher={American Society of Mechanical Engineers}
}

@article{jin2024mare,
  title={Mare: Multi-agents collaboration framework for requirements engineering},
  author={Jin, Dongming and Jin, Zhi and Chen, Xiaohong and Wang, Chunhui},
  journal={arXiv preprint arXiv:2405.03256},
  year={2024}
}

@article{yang2025docagent,
  title={DocAgent: A Multi-Agent System for Automated Code Documentation Generation},
  author={Yang, Dayu and Simoulin, Antoine and Qian, Xin and Liu, Xiaoyi and Cao, Yuwei and Teng, Zhaopu and Yang, Grey},
  journal={arXiv preprint arXiv:2504.08725},
  year={2025}
}

@article{luo2024repoagent,
  title={Repoagent: An llm-powered open-source framework for repository-level code documentation generation},
  author={Luo, Qinyu and Ye, Yining and Liang, Shihao and Zhang, Zhong and Qin, Yujia and Lu, Yaxi and Wu, Yesai and Cong, Xin and Lin, Yankai and Zhang, Yingli and others},
  journal={arXiv preprint arXiv:2402.16667},
  year={2024}
}

@article{yuan2024evaluating,
  title={Evaluating and improving chatgpt for unit test generation},
  author={Yuan, Zhiqiang and Liu, Mingwei and Ding, Shiji and Wang, Kaixin and Chen, Yixuan and Peng, Xin and Lou, Yiling},
  journal={Proceedings of the ACM on Software Engineering},
  volume={1},
  number={FSE},
  pages={1703--1726},
  year={2024},
  publisher={ACM New York, NY, USA}
}

@inproceedings{deng2024pentestgpt,
  title={$\{$PentestGPT$\}$: Evaluating and harnessing large language models for automated penetration testing},
  author={Deng, Gelei and Liu, Yi and Mayoral-Vilches, V{\'\i}ctor and Liu, Peng and Li, Yuekang and Xu, Yuan and Zhang, Tianwei and Liu, Yang and Pinzger, Martin and Rass, Stefan},
  booktitle={33rd USENIX Security Symposium (USENIX Security 24)},
  pages={847--864},
  year={2024}
}

\end{document}